\documentclass[10pt,a4paper]{article}
\usepackage{jheppub_kim}
\usepackage{pdflscape}
\usepackage{amsmath}
\usepackage{amssymb}
\usepackage{dcolumn}
\usepackage{bm}
\usepackage{color}
\usepackage{amsmath}
\usepackage{breqn}
\usepackage{graphicx}
\usepackage{epsfig}
\usepackage{caption,subcaption}
\usepackage{amsfonts}
\usepackage{graphicx}
\usepackage{parskip}
\usepackage{dcolumn}
\UseRawInputEncoding

\begin{document}

\title{Interplay of Holographic and New Agegraphic Dark Energy in Cosmology: A hybrid dark energy model from a generalized length-time cut-off}

\author[a]{Sayani Maity}
\author[b]{Aritra Sanyal}
\author[c]{Prabir Rudra}

\affiliation[a] {Department of Mathematics, Sister Nivedita University, DG-1/2, Action Area 1, New Town, Kolkata 700 156, India}

\affiliation[b] {Department of Mathematics, Jadavpur University, Kolkata-700 032, India.}

\affiliation[c] {Department of Mathematics, Asutosh College,
Kolkata-700 026, India}

\emailAdd{sayani.office88@gmail.com}
\emailAdd{aritrasanyal1@gmail.com}
\emailAdd{prudra.math@gmail.com}

\abstract{The spatial and temporal infrared cutoffs are the source of the Holographic dark energy (HDE) and New Agegraphic dark energy (NADE) models, respectively. Inspired by the spacetime unification of space and time in General Relativity, we propose that a combined spacetime cutoff should govern the dark energy density, with HDE and NADE appearing as limiting cases when the spatial or temporal contribution predominates. In this connection, we explore a hybrid model of holographic and new Agegraphic dark energy, where the energy density of the resulting model is a combination of the two models. We consider an interacting and a non-interacting scenario between the hybrid dark energy model and cold dark matter. Cosmological implications of the model is studied via different cosmological parameters like the equation of state parameter, deceleration parameter, statefinder parameter, and Om-diagnostic. A stability check for the model has been performed using the squared speed of sound. Finally the parameter space of the model is constrained using observational data like Hubble data, BAO data and DESI data. We have also checked the Hubble tension for our hybrid model and found it to be substantially low in comparison to other models. From our analysis we see that the constructed hybrid dark energy model can describe the evolution of the universe successfully.}

\maketitle

\section{Introduction}
One of the most significant challenges of contemporary cosmology is the discovery of the late-time accelerated expansion of the Universe through the observations of Type Ia supernovae \cite{perl1, perl2, perl3}. Large-scale structural surveys, baryon acoustic oscillations \cite{bao1, bao2}, and cosmic microwave background radiation \cite{cmb1, cmb2} studies have all corroborated this accelerated phase. According to General Relativity (GR), the acceleration is typically explained by dark energy (DE) \cite{brax}, an exotic component that makes up almost $70\%$ of the universe's total energy. The actual nature of DE remains one of the most profound challenges in modern cosmology. While the cosmological constant $\Lambda$, characterized by a constant equation of state $\omega=-1$, provides an excellent phenomenological fit to current observations within the $\Lambda$CDM paradigm, it suffers from severe theoretical difficulties, most notably the fine-tuning problem and cosmic coincidence problem \cite{Peebles2003, Padmanabhan2003, Copeland2006, Frieman2008, Caldwell2009, Silvestri2009, Li2011,Bamba2012, Li2013}. These shortcomings have motivated the exploration of dynamical dark energy scenarios in which both the energy density and equation of state evolve with cosmic time \cite{10}. Among such approaches, holographic dark energy (HDE), inspired by the holographic principle, and new agegraphic dark energy (NADE), rooted in quantum fluctuations of spacetime, have attracted considerable attention due to their firm connections with fundamental physics and their foundations in quantum gravitational principles
\cite{11_Nojiri_2006,12_Cai_2007,13_Wei_2008,14_Wei_2008,17_Zhang_2013}. These frameworks offer promising avenues to address longstanding cosmological issues such as the cosmological constant problem (fine tuning) and the cosmic coincidence problem. 

The holographic concept provides a connection between ultraviolet and infrared cutoffs and results in a dark energy density that is based on the universe's infrared length scale. Numerous studies have demonstrated that HDE models offer a plausible explanation for the observed cosmic acceleration. The choice of the infrared cutoff, however, has a significant impact on the cosmic history of HDE and may provide theoretical and phenomenological difficulties.
In \cite{ASSMUDAP25} the influence of different infrared cut-offs on generalized entropy-based holographic dark energy was explored. By considering various horizon choices such as Tsallis, R\'enyi and Sharma--Mittal, the Holographic Dark Energy model has been investigated within the framework of Ho\v{r}ava--Lifshitz gravity. The authors demonstrated that the cosmological evolution is highly sensitive to the choice of the infrared cut-off as well as the underlying entropy formalism. Other notable work in holographic dark energy can be found in \cite{hh1, hh2, hh3, hh4, hh5}. The original Agegraphic dark energy (ADE) model suffers from several conceptual and phenomenological shortcomings. In particular, it fails to naturally reproduce the standard sequence of cosmic evolution, predicts an improper scaling of dark energy with the scale factor, and yields an equation of state parameter that remains too close to zero at early times, leading to tensions with observational data and structure formation. Moreover, defining the characteristic time scale as the total age of the universe renders the model intrinsically non-local in time, since the energy density depends on the entire past history of cosmic evolution. To overcome these limitations, the new agegraphic dark energy (NADE) model \cite{Neupane:2007ra} was proposed by replacing the age of the universe with the conformal time. Specifically, quantum fluctuations of spacetime resulting from the Károlyłazy uncertainty relation are the foundation of the NADE paradigm. This modification significantly improves the dynamical behavior of the model, allowing a natural alleviation of the coincidence problem while preserving a minimal parameter space comparable to that of the $\Lambda$CDM model. Owing to its economy and improved theoretical consistency, the NADE model has been shown to be in good agreement with current observational constraints. NADE offers a dynamical dark energy framework that can describe the evolution of the Universe without introducing an explicit cosmological constant. However, NADE is limited by increasingly accurate cosmic measurements, just like other dark energy models. Recently, in \cite{SMASPR26} we have studied the NADE model in the background of loop quantum cosmology.
In recent years, several generalized entropy formalisms have been incorporated into these models, leading to a variety of modified dark energy scenarios with improved phenomenological features. The introduction of generalized entropy measures, such as Barrow, Tsallis, R\'enyi and Sharma--Mittal entropies, has significantly broadened the scope of holographic and agegraphic dark energy models. These generalized entropies account for possible quantum gravitational corrections to the Bekenstein--Hawking entropy and consequently modify the corresponding dark energy density. In \cite{SMUD19}, Tsallis, Rényi and Sharma-Mittal holographic and new agegraphic dark energy models have been investigated in the context of D-dimensional fractal universe. In ref \cite{PSSMUD22}, the authors reconstructed extended cubic gravity using both entropy-corrected holographic dark energy and entropy-corrected new agegraphic dark energy. Their results illustrated that different quantum entropy corrections can naturally generate viable modified gravity theories capable of explaining the observed cosmic acceleration without introducing additional exotic matter components. A systematic reconstruction of modified gravity theories from generalized agegraphic dark energy was carried out in the work entitled \cite{SMAK25}, where the authors reconstructed viable $f(R)$ gravity models corresponding to both Barrow agegraphic dark energy (BADE) and new Barrow agegraphic dark energy (NBADE). 

The holographic relationship between the ultraviolet and infrared sectors of an effective field theory is reflected in the HDE scenario, where the dark energy density is defined by an infrared cutoff with a cosmic length scale. The NADE model, on the other hand, links the dark energy density to a cosmological time scale, specifically the universe's conformal age, and is based on the Karolyhazy uncertainty relation. Both of these theories depend on the existence of a fundamental infrared cutoff controlling the vacuum energy density, although coming from different physical considerations. According to General Relativity, space and time are combined into a four-dimensional spacetime manifold rather than existing as separate entities. It is reasonable to anticipate that a genuine description of dark energy should not rely solely on either a length scale or a time scale as spatial and temporal coordinates are complimentary parts of the same geometrical structure. Alternatively, the effective dark energy density may be influenced concurrently by temporal and spatial infrared cutoffs. A hybrid dark energy model that combines the key elements of both holographic and agegraphic techniques may be built, according to this point of view. Therefore, the suggested hybrid framework can be thought of as a more general implementation of the infrared cutoff concept, in which the growth of dark energy is simultaneously determined by the Universe's distinctive length and time scales. In addition to combining two highly motivated classes of dark energy models into a single theoretical framework, this construction enables one to examine whether the combined effects of temporal and spatial cutoffs can improve the description of cosmic acceleration and mitigate some of the drawbacks of individual models. Moreover, as HDE and NADE originate from distinct facets of spacetime physics, their combination could provide more profound understanding of the relationship between holography, quantum gravity, and the nature of dark energy.

Motivated by these considerations, in the present work we propose a hybrid dark energy model in which the total dark energy density is taken as the sum of the holographic dark energy density and the new agegraphic dark energy density. By examining a number of significant cosmological factors, such as the equation of state parameter, deceleration parameter, statefinder diagnostics, and Om diagnostic, we aim to explore the cosmological implications of this model. These quantities enable comparisons with conventional dark energy scenarios and offer important insights into the evolutionary behavior of the model. We would also like to check the stability of the model to assess the model's physical feasibility. Apart from this, we would like to carry out a thorough parameter estimation using current cosmological datasets in order to evaluate the observational consistency of the suggested framework. The generated constraints enable us to assess the hybrid model's capacity to explain the available observational data and identify the admissible parameter space. Our goal is to investigate whether a dark energy scenario that combines holographic and agegraphic contributions can offer a workable and empirically validated substitute for traditional dark energy models.

The paper is organized as follows: In section 2 we propose the hybrid dark energy model, with the background gravity as GR. We frame the set-up for both the interacting and non-interacting cases between DE and cold dark matter. In section 3 we explore the cosmology of the hybrid DE model. Section 4 is dedicated to constraining the parameter space of the DE model. In section 5, we perform a Hubble tension analysis for our hybrid dark energy model. Finally the paper ends with a discussion and conclusion in section 6.

\section{Hybrid Dark energy Model}
Within the framework of General Relativity (GR), the Friedmann-Robertson-Walker (FRW) equations describe the dynamics of a homogeneous and isotropic universe. These equations are expressed as

\begin{equation}\label{FRW1}
H^2 = \frac{8\pi G}{3}\rho - \frac{k}{a^2}    
\end{equation}
and
\begin{equation}\label{FRW2}
\quad \frac{\ddot{a}}{a} = -\frac{4\pi G}{3}(\rho + 3p)
\end{equation}
where $H = \frac{\dot{a}}{a}$ denotes the Hubble parameter and $a(t)$ is the cosmological scale factor. Here, $G$ is Newton’s gravitational constant, and $k$ represents the spatial curvature of the universe. The cases $k = 0, 1, -1$ correspond to flat, closed, and open universes, respectively. The symbols $\rho$ and $p$ denote the total energy density and pressure of cosmic matter. Currently, the dominant components of the universe are dark energy (DE) and dark matter (DM). The accelerated expansion of the universe is primarily attributed to DE, making it a crucial subject of investigation. Thus, the total energy density and pressure can be written as~ $\rho = \rho_{D} + \rho_m$~ and~ $p = p_{D} + p_m$,~ considering only the dominant components. Since the pressure of matter is negligible $p_m \approx 0$, it follows that $p = p_{D}$.\\
Here we consider a flat homogeneous and isotropic FRW universe described by the following field equations:

\begin{equation}\label{FRW3}
H^2 = \frac{8\pi G}{3}\rho    
\end{equation}

\begin{equation}\label{FRW4}
\quad \frac{\ddot{a}}{a} = -\frac{4\pi G}{3}(\rho + 3p)   
\end{equation}

Moreover, we have the continuity equation of the energy and matter components given by
\begin{equation}\label{FRW5}
\dot{ \rho}+3H (\rho+p)=0
\end{equation}

The study of black hole thermodynamics led Gerard 't Hooft to propose the holographic principle (HP), which suggests that all information within a spatial volume can be described by a theory defined on its boundary. Leonard Susskind later provided a precise string-theoretic formulation of this principle, and Maldacena's celebrated AdS/CFT correspondence offered its most successful realization. In recent days, HP is widely regarded as a cornerstone of quantum gravity, with applications across diverse fields. In nuclear physics, the AdS/QCD framework has been used to probe quark-gluon plasma. In condensed matter physics, AdS/CMT has illuminated superconductivity and super fluidity. In theoretical physics, HP has inspired the concept of holographic entanglement entropy, and in cosmology, it has been applied to the study of de Sitter space and inflation. Given that DE may ultimately be a quantum gravity phenomenon, HP offers a promising avenue for resolution. In 2004, Miao Li introduced the holographic dark energy (HDE) model, the first DE framework explicitly inspired by HP. Remarkably, HDE not only embodies the holographic principle but also aligns well with current cosmological observations, making it a strong candidate for explaining dark energy. The energy density of holographic dark energy model is given by \cite{LI20041}
\begin{equation}\label{NON1}
\rho_{HDE} = 3c^2 M_P^2 L^{-2}  
\end{equation}
where $L$ is the length scale, $M_P$ is the reduced Planck mass, and $c$ is a numerical constant introduced for convenience. We consider HDE with Hubble horizon cut-off $L=1/H$.

The Agegraphic dark energy (ADE) idea starts from quantum fluctuations of spacetime. Using the Karolyhazy relation together with the time-energy uncertainty one estimates an energy density that scales with an inverse square of a cosmological time scale. In the original ADE proposal the relevant time is the cosmic age $T$ giving the energy density
$\rho_{ADE}=3n^{2}M_P^{2}T^{-2}$ where $n$ is a dimensionless parameter that absorbs uncertainties. ADE naturally yields a dark energy component whose present magnitude can be of the observed order $n \sim O(1)$, and it avoids the causality issue that affects some holographic constructions based on the future event horizon. However, ADE exhibits a tracking problem in the early universe. When $T$ is taken as the age, the model can track the dominant component (matter or radiation) and fail to evolve into a late-time dark energy dominated phase without additional ad-hoc modifications. To cure this, the new agegraphic dark energy (NADE) replaces the cosmic age by the conformal time $\eta$. The energy density of NADE model is given by \cite{Wei:2007ty}  
\begin{equation}\label{NON1}
\rho_{NADE} = 3n^2M_{p}^2\eta^{-2}  
\end{equation}
where the conformal time $\eta$ is given by
\begin{equation}\label{NON2}
\eta=\int_0^t \frac{dt}{a}=\int_0^a \frac{da}{Ha^2}
\end{equation}

Since our objective is to construct a hybrid HDE-NADE dark energy model based on the idea that length and time are complementary manifestations of spacetime, then the key step is to define a generalized infrared cutoff that reduces to the HDE cutoff in one limit and the NADE cutoff in another. We have seen that both the HDE and NADE densities scale as the inverse square of a fundamental scale. So a natural spacetime cutoff $\mathcal{L}$ should satisfy $\rho_{D}=3M_{P}^{2}\mathcal{L}^{-2}$. There can be various ways in which this natural spacetime cut-off can be constructed. We will follow the most obvious path of harmonic mean cut-off where $\mathcal{L}$ is given by
\begin{equation}\label{cut}
\frac{1}{\mathcal{L}^{2}}=\lambda\frac{1}{L^2}+\gamma\frac{1}{\eta^{2}}
\end{equation}
where $\lambda$ and $\gamma$ are constants.

Here, we will adopt a power-law form for the scale factor as presented in \cite{21_PhysRevD.74.086009}. It is given by
\begin{equation}\label{NON3}
a(t)=a_0 t^m, ~~m>0
\end{equation}
where $a_0$ and $m$ are the current value of the scale factor and the power-law exponent, respectively. This form of the scale factor implies an initial Big Bang singularity but does not lead to any finite-time future singularities. Various studies have investigated the cosmological implications of such scale factors in the context of different modified gravity theories. For this form of scale factor, the energy density of NADE and HDE models take the form,
\begin{equation}\label{NON4}
\rho_{NADE}=3n^2M_{P}^2(1-m)^{2}a_0^{\frac{2}{m}} (1+z)^{\frac{2}{m}-2}
\end{equation}
\begin{equation}
\rho_{HDE}=3c^2 M_P^2 m^2 a_0^{\frac{2}{m}}(1+z)^{\frac{2}{m}}
\end{equation}
where $z$ represents the cosmological redshift, given by $z=\frac{1}{a}-1$.
Using the generalized cut-off constructed in eqn.(\ref{cut}), we propose a hybrid dark energy model combining the features of HDE and NADE model whose energy density is given by
\begin{equation}
\rho_D=\alpha \rho_{NADE}+\beta \rho_{HDE}, ~~~~ \alpha, \beta\neq 0
\end{equation}
For $\alpha=0$, we get the pure HDE model, whereas for $\beta=0$, we get the the pure NADE model. The above hybrid density is not merely a mathematical sum, but it has a strong physical motivation. HDE and NADE correspond to two different infrared manifestations of quantum gravitational effects - one associated with holographic entropy bounds and the other with spacetime quantum fluctuations. Thus, the hybrid model investigates the potential that both mechanisms contribute to the effective dark energy sector at the same time. For power-law type of scale factor $\rho_D$ takes the form
\begin{equation}\label{DE Expression}
\rho_D=3 M_P^2a_0^{\frac{2}{m}}(1+z)^{\frac{2}{m}}\Big{[}\alpha n^2(1-m)^2(1+z)^{-2}+\beta c^2 m^2\Big{]}
\end{equation}
Here we will consider Plank units where $M_P=1$. In the next subsection we will consider two different cases of interaction between the DE and DM sectors of the universe.

\subsection{Non-interacting Case}
If we consider that there is no interaction between dark energy and cold dark matter and both are conserved separately, the conservation equation (\ref{FRW3}) yields
\begin{equation}\label{NON5}
\dot{\rho}_{m}+3H \rho_m=0,
\end{equation}
and
\begin{equation}\label{NON6}
\dot{\rho}_{D}+3H \rho_D (1+\omega_D)=0.
\end{equation}
where $\omega_D=p_{D}/\rho_{D}$ is the equation of state (EoS) parameter of the hybrid DE. Solving equation (\ref{NON5}) we get the matter density as,
\begin{equation}\label{NON7}
\rho_{m}=\rho_{m_0}(1+z)^3
\end{equation}
where $\rho_{m_0}$ is the present value of the matter density. Using the above equations the expression for the Hubble parameter $H(z)$ for the universe filled with hybrid DE and cold dark matter is obtained as,
\begin{equation}
H(z)^2=\frac{M_P^2 (1+z)^3}{1-\beta c^2 M_P^2}\left[ \alpha  n^2  a_0^{\frac{2}{m}} (1-m)^2 (1+z)^{-5+\frac{2}{m}}+\Omega_{m0} H_0^2\right]
\end{equation}
where $\Omega_{m0}=\rho_{m_0}/3M_{P}^{2}H_{0}^{2}$ is the dimensionless matter density parameter, and $H_0$ is the present value of the Hubble parameter.
Differentiating equation (\ref{FRW2}) with respect to cosmic time $t$ and exploiting equations (\ref{DE Expression}) and (\ref{NON7}) we get
\begin{equation}
\frac{\dot{H}}{H^2}=\frac{a_0^{\frac{1}{m}} M_P^2 (1+z)^{3+\frac{1}{m}}}{2H^3} \Big{[}2 a_0^{\frac{2}{m}} (1+z)^{\frac{2}{m}-1} \left\{\alpha n^2 (m-1)^3 -\beta c^2 m^2 (1+z)^2\right\} -3m H_0^2 \Omega_{m0}\Big{]}
\end{equation}

\subsection{Interacting Case}
Now we consider an interaction between hybrid DE and cold DM, governed by the continuity equations,
\begin{equation}\label{IN1}
\dot{ \rho_{m}}+3H \rho_m= Q
\end{equation}
and
\begin{equation}\label{IN2}
\dot{ \rho_{D}}+3H \rho_D (1+\omega_D)=-Q
\end{equation}

Here, $Q$ is a dynamic quantity that acts as the interaction term between Cold Dark Matter (CDM) and the dark energy. Typically, three commonly used forms of $Q$ are considered in literature,
$$
Q_1 = 3 \delta H \rho_D, \quad Q_2 = 3 \delta H (\rho_m + \rho_D), \quad Q_3 = 3 \delta H \rho_m,
$$

where $\delta$ is the interaction constant representing the strength of the energy exchange between CDM and DE. 
In this work, we consider the third interaction form, $Q_3$, which is extensively used in the literature to describe energy exchange between DE and CDM under different cosmological conditions. The direction of energy flow depends on the sign of the coupling constant $\delta$. A positive $\delta$ signifies that DE decays into CDM, while a negative $\delta$ indicates a transfer of energy from CDM to DE. Both observational evidence and theoretical analyses predominantly support the case where DE decays into CDM, i.e., $\delta>0$.

Solving equation (\ref{IN1}) for $Q_3$ type of interaction, we get,
\begin{equation}\label{IN3}
\rho_m=\rho_{m0} (1+z)^{3-3\delta}
\end{equation}
Using the above equations the expression for $H(z)$ is obtained as,
\begin{equation}
H(z)^2=\frac{M_P^2 (1+z)^{3-3\delta}}{1-\beta c^2 M_P^2}\left[ \alpha  n^2  a_0^{\frac{2}{m}} (1-m)^2 (1+z)^{3\delta-5+\frac{2}{m}}+\Omega_{m0} H_0^2\right]
\end{equation}
Differentiating equation (\ref{FRW2}) with respect to cosmic time $t$ and exploiting equations (\ref{DE Expression}) and (\ref{IN3}) we get
\begin{equation}
\frac{\dot{H}}{H^2}=\frac{a_0^{\frac{1}{m}} M_P^2 (1+z)^{2+\frac{3}{m}}}{2H^3} \Big{[}2 a_0^{\frac{2}{m}}  \left\{\alpha n^2 (m-1)^3 -\beta c^2 m^2 (1+z)^2\right\} +3m (\delta-1) H_0^2 \Omega_{m0}(1+z)^{1-3\delta-\frac{2}{m}}\Big{]}
\end{equation}

\section{Cosmology of Hybrid DE model}
In this section, we will explore the cosmological implications of the proposed hybrid DE model. Various cosmological parameters such as the equation of state parameter, deceleration parameter, statefinder parameter, Om diagnostic, etc. will be studied to probe the cosmological competence of the proposed hybrid DE model.

\subsection{Equation of state (EoS) and Deceleration Parameter}
Here, we examine the cosmological consequences of the equation of state (EoS) parameter, defined as
\begin{equation}\label{EoS0}
\omega_{d}=\frac{p_{d}}{\rho_{d}},
\end{equation}
where $\rho_{d}$ and $p_{d}$ denote the energy density and pressure of the dark energy component, respectively. The value of $\omega_{d}$ plays a crucial role in characterizing the dynamical behavior of the universe. In particular, if $\omega_{d}$ crosses the phantom divide line $\omega_{d}=-1$ from $\omega_{d}>-1$ to $\omega_{d}<-1$, the corresponding evolution is referred to as \emph{quintom} behavior. The different cosmological phases along with the corresponding values of the EoS parameter is given in the following table.

\begin{table}[ht]
\centering
\begin{tabular}{|c|c|}
\hline
Value of $\omega_{d}$ & Cosmological Phase \\ [0.5ex]
\hline
$\omega_{d}=0$ & Non-relativistic (dust) matter \\
$-1<\omega_{d}<-\frac{1}{3}$ & Quintessence regime \\
$\omega_{d}=-1$ & Cosmological constant ($\Lambda$CDM) \\
$\omega_{d}<-1$ & Phantom regime \\
\hline
\end{tabular}
\end{table}

\subsubsection{Non-interacting Case}
Exploiting equations (\ref{FRW3}), (\ref{FRW4}), (\ref{DE Expression}), (\ref{NON6}), (\ref{NON7}) and (\ref{EoS0}) we get the expression for the EoS parameter as
\begin{equation}\label{omegaNONINT}
\omega_D=-\frac{1}{3}-\frac{1}{M_P^2}\Big{[}\frac{6m(m-1)+3H_0^2 M_P^2 a_0^{-\frac{2}{m}}\Omega_{m0}(1+z)^{3-\frac{2}{m}}}{\alpha n^2 (1-m)^2(1+z)^{-2}+\beta c^2 m^2 }\Big{]}
\end{equation}

\begin{figure}[hbt!]
\begin{center}
\includegraphics[height=2.5in]{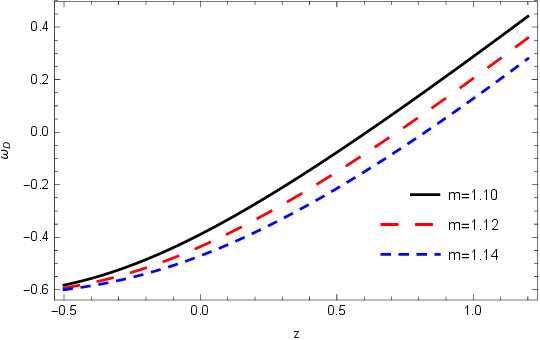}\\
\vspace{3mm}
\caption{The figure shows the plot of EoS parameter $\omega_{D}$ for the hybrid model against redshift for the non-interacting case for three different values of $m$. Other parameters are considered as $H_0=69$, $\Omega_{m0}=0.29$, $a_0=1.2952$, $\alpha=1.009$, $\beta=0.07$ and $n=2.5$. }
\label{figscale11}
\end{center}
\end{figure}

\begin{figure}[hbt!]
\begin{center}
\includegraphics[height=2.5in]{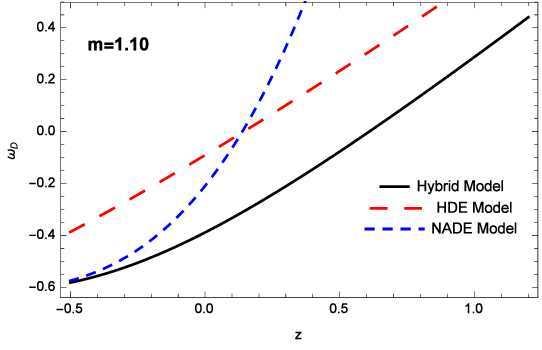}\\
\vspace{3mm}
\caption{The figure compare the plot of EoS parameter $\omega_{D}$ for Hybrid, HDE ($\alpha=0$) and NADE  ($\beta=0$) against redshift for the non-interacting case for $m=1.10$  }
\label{figscale11a}
\end{center}
\end{figure}

\begin{figure}[hbt!]
\begin{center}
\includegraphics[height=2.5in]{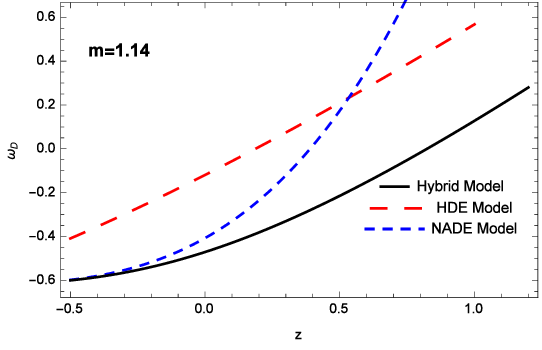}\\
\vspace{3mm}
\caption{The figure compare the plot of EoS parameter $\omega_{D}$ for Hybrid, HDE ($\alpha=0$) and NADE  ($\beta=0$) against redshift for the non-interacting case for $m=1.14$ }
\label{figscale11b}
\end{center}
\end{figure}

The plot of $\omega_D$ for the non-interacting case is shown in figure (\ref{figscale11}) for different values of the power-law parameter $m$. We observe that for all these values of $m$ the model yields quintessence-like behavior at present time ($z=0$). This is because we see that around $z=0$ we have $-1<\omega_{D}<-1/3$. Moreover, we see that in the early times there was a domination of matter and the effect of DE was negligible. In the future ($z<0$), the trajectories tend to remain in the quintessence regime. It is also evident that the trajectories shift towards more negative values as the value of $m$ increases. In figures (\ref{figscale11a}) and (\ref{figscale11b}) we have generated the comparative $\omega_{D}$ vs $z$ plots for two different values of $m$. In these plots we have compared $\omega_{D}$ trajectories for the hybrid model, pure HDE model, and the pure NADE model. In both the plots we see that the EoS for the hybrid model stands out in comparison with the other two models. The trajectory for the hybrid model prominently lies in the proper quintessence regime in the present time ($z=0$) in comparison to the other two models.

The deceleration parameter, represented by $q$, is a dimensionless number that characterizes the rate of change of cosmic expansion. It shows whether the universe's expansion is quickening or slowing down. It is defined as
\begin{equation}
q=-1-\frac{\dot{H}}{H^2}
\end{equation}
$q>0$ indicates a decelerating universe and $q<0$ presents
an accelerating phase of the universe. Using the above equations the deceleration parameter for the non-interacting case is obtained as,
\begin{equation}
q=-1-\frac{ M_P^2 (1+z)^{3-\frac{2}{m}}}{2m^3a_0^{\frac{2}{m}}} \Big{[}2 a_0^{\frac{2}{m}} (1+z)^{\frac{2}{m}-1} \left\{\alpha n^2 (m-1)^3 -\beta c^2 m^2 (1+z)^2\right\} -3m H_0^2 \Omega_{m0}\Big{]}
\end{equation}

\begin{figure}[hbt!]
\begin{center}
\includegraphics[height=2.5in]{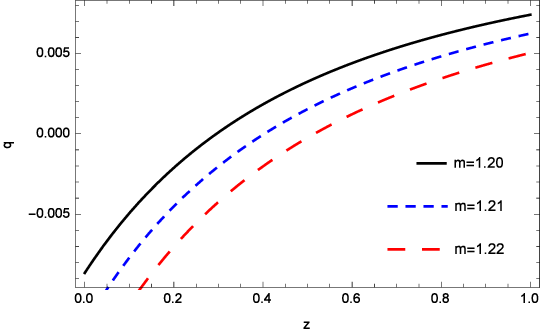}\\
\vspace{3mm}
\caption{The figure shows the plot of deceleration parameter $q$ for the hybrid model against redshift for the non-interacting case. Other parameters are considered as $H_0=69$, $\Omega_{m0}=0.29$, $a_0=1.2952$, $\alpha=0.74$, $\beta=0.12$ and $n=2.5$.}
\label{figscale12}
\end{center}
\end{figure}

The plot of $q$ for the non-interacting case is shown in figure (\ref{figscale12}) for different values of the power-law parameter $m$. We observe that for all values of $m$ the trajectories undergo a transition from the decelerating ($q>0$) to an accelerating phase ($q<0$) around $z=0.4$ which is observationally favored.

\subsubsection{Interacting Case}
Exploiting equations (\ref{FRW3}), (\ref{FRW4}), (\ref{DE Expression}), (\ref{IN1}) and (\ref{EoS0}) we get the EoS parameter for the interacting case as,
\begin{equation}\label{omegaINT}
\omega_D=-\frac{1}{3}-\frac{1}{3M_P^2}\Big{[}\frac{2m(m-1)+H_0^2 M_P^2 a_0^{-\frac{2}{m}}\Omega_{m0}(1+z)^{3-\frac{2}{m}-3\delta}}{\alpha n^2 (1-m)^2(1+z)^{-2}+\beta c^2 m^2 }\Big{]}
\end{equation}

\begin{figure}[hbt!]
\begin{center}
\includegraphics[height=2.5in]{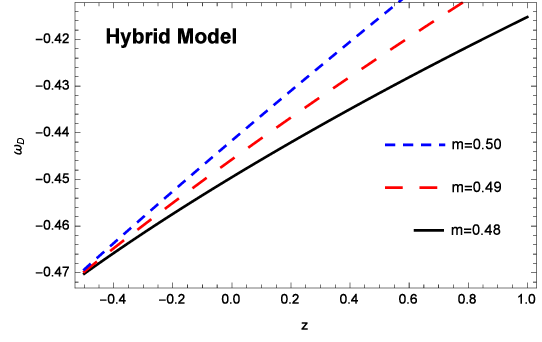}\\
\vspace{3mm}
\caption{The figure shows the plot of EoS parameter $\omega_{D}$ for the hybrid model against redshift for the interacting case for three different values of $m$. Other parameters are considered as $H_0=69$, $\Omega_{m0}=0.29$, $a_0=$, $\alpha=$, $\beta=$ and $n=2.5$}
\label{figscale21}
\end{center}
\end{figure}

\begin{figure}[hbt!]
\begin{center}
\includegraphics[height=2.5in]{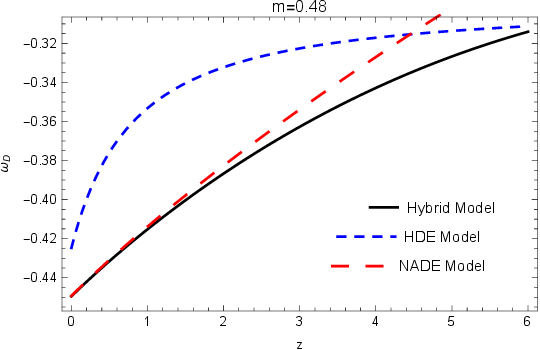}\\
\vspace{3mm}
\caption{The figure compare the plot of EoS parameter $\omega_{D}$ for Hybrid, HDE ($\alpha=0$) and NADE  ($\beta=0$) against redshift for the interacting case for $m=0.48$}
\label{figscale21a}
\end{center}
\end{figure}

\begin{figure}[hbt!]
\begin{center}
\includegraphics[height=2.5in]{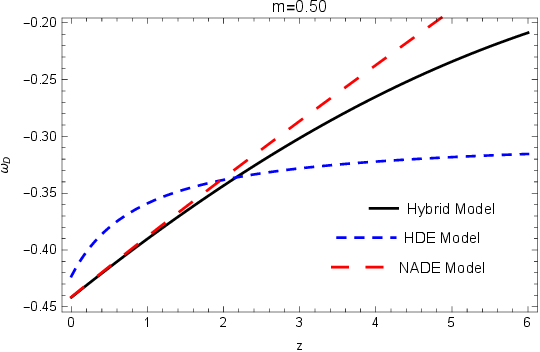}\\
\vspace{3mm}
\caption{The figure compare the plot of EoS parameter $\omega_{D}$ for Hybrid, HDE ($\alpha=0$) and NADE  ($\beta=0$) against redshift for the interacting case for $m=0.50$}
\label{figscale21b}
\end{center}
\end{figure}

The plot of $\omega_D$ for the interacting case is shown in figure (\ref{figscale21}) for three different values of the power-law parameter $m$. We observe that for all the three values of $m$ the model presents quintessence-type behavior in recent times ($z\approx 0$). Comparing figs.(\ref{figscale11}) and (\ref{figscale21}) we see that with the introduction of interaction between DE and CDM the trajectories change, showing the effect of interaction. In figures (\ref{figscale21a}) and (\ref{figscale21b}) we have generated EoS parameter plots in a comparative scenario between the hybrid model, HDE model and the NADE model for the interacting case. Clear distinction in the respective trajectories are found. In the present time ($z=0$), the $\omega_{D}$ value is least for the hybrid model showing greater exotic nature of the same, compared to HDE and NADE. This shows enhanced ability for the hybrid model to drive the late cosmic acceleration.

The expression for the deceleration parameter for the interacting case is obtained as,
\begin{equation}
q =-1-\frac{ M_P^2 (1+z)^{2}}{2m^3a_0^{\frac{2}{m}}} \Big{[}2 a_0^{\frac{2}{m}}  \left\{\alpha n^2 (m-1)^3 -\beta c^2 m^2 (1+z)^2\right\} +3m (\delta-1) H_0^2 \Omega_{m0}(1+z)^{1-3\delta-\frac{2}{m}}\Big{]}
\end{equation}

\begin{figure}[hbt!]
\begin{center}
\includegraphics[height=2.5in]{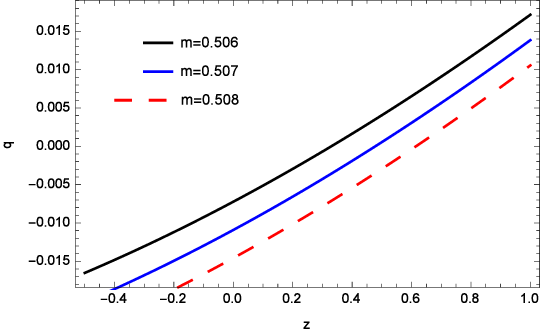}\\
\vspace{3mm}
\caption{The figure shows the plot of deceleration parameter $q$ for the hybrid model against redshift for the interacting case. Other parameters are considered as $H_0=69$, $\Omega_{m0}=0.29$, $a_0=1.2952$,$\delta=0.2$ $\alpha=0.991$, $\beta=0.0508$ and $n=2.5$.}
\label{figscale22}
\end{center}
\end{figure}
The plot of $q$ for the interacting case is shown in Figure (\ref{figscale22}) for three different values of the power-law parameter $m$. In the plot we see that just like the non-interacting case, here also we have a transition from the decelerated regime to an accelerated regime around $z\approx 0.45$, which is favored observationally. 

\subsection{Om diagnostic}
The Om diagnostic is a purely geometric tool constructed from the Hubble parameter and the redshift. It is designed to distinguish evolving dark-energy scenarios from the standard $\Lambda$CDM model, either with or without explicitly specifying the matter density. A constant $Om(z)$ across redshift indicates that dark energy behaves as a cosmological constant. In contrast, a positive redshift dependence of $Om(z)$, i.e., $d~Om(z)/dz>0$ signals phantom behavior, corresponding to an EoS parameter $\omega_D < -1$. A negative redshift dependence of $Om(z)$, i.e.,  $d~Om(z)/dz<0$ indicates a quintessence-like behavior corresponding to $-1<\omega_{D}<-1/3$.

For a spatially flat universe, $Om(z)$ is defined as \cite{Zunckel2008, Sahni2008} 
\begin{equation}
Om(z)=\frac{\frac{H^2(z)}{H_0^2}-1}{(1+z)^3-1},
\end{equation}
where $H_0$ is the present value of the Hubble parameter. For a dark energy component with constant equation of state parameter $\omega_D$ we have the $Om(z)$ as
\begin{equation}
Om(z)=\Omega_{m0}+(1- \Omega_{m0})\frac{(1+z)^{3(1+\omega_D)}-1}{(1+z)^3-1} 
\end{equation}
This expression shows that $\Lambda$CDM yields $Om(z)=\Omega_{m0}$, while quintessence models give $Om(z)>\Omega_{m0}$ and phantom models give $Om(z)<\Omega_{m0}$. Importantly, the $Om$ diagnostic serves as a null test for $\Lambda$CDM model. Moreover, an increasing $Om(z)$ at late times implies a decaying or evolving dark energy model \cite{Shafieloo2009}.

\begin{figure}[hbt!]
\begin{center}
\includegraphics[height=2.5in]{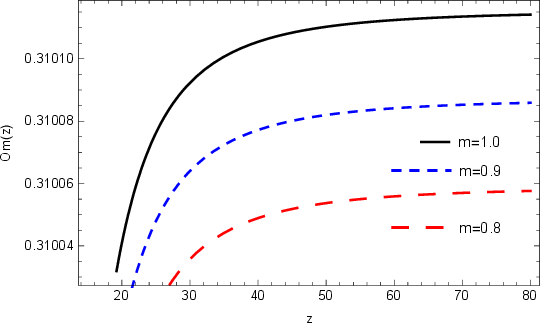}\\
\vspace{3mm}
\caption{The figure shows the plot of $Om$ diagnostic for the hybrid model against redshift for the non-interacting case. Other parameters are considered as $H_0=69$, $\Omega_{m0}=0.29$, $a_0=1.2952$, $\alpha=0.94$, $\beta=0.1123$ and $n=2.5$.}
\label{figscale71}
\end{center}
\end{figure}

\begin{figure}[hbt!]
\begin{center}
\includegraphics[height=2.5in]{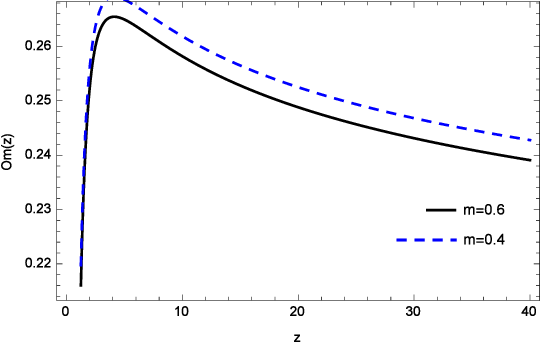}\\
\vspace{3mm}
\caption{The figure shows the plot of $Om$ diagnostic for the hybrid model against redshift for the interacting case. Other parameters are considered as $H_0=69$, $\Omega_{m0}=0.29$, $a_0=1.502$, $\delta=0.02$, $\alpha=0.991$, $\beta=0.0508$ and $n=2.5$.}
\label{figscale72}
\end{center}
\end{figure}

$Om(z)$ is plotted for three different values of $m$ both in non-interacting and interacting cases in figures (\ref{figscale71}) and (\ref{figscale72}) respectively. For non-interacting case the trajectories of $Om(z)$ reflects quintessence like behavior as we approach towards $z=0$ ($Om(z)$ decreases as $z$ decreases). For the interacting scenario, we see that initially the trajectories rise ($Om(z)$ increases) as $z$ decreases, showing a phantom-like behavior. But as we approach $z=0$, the trajectories dive deep, showing a quintessence-like behavior eventually. 


\subsection{Cosmological Planes}
Nowadays, one of the most effective tools for distinguishing different dark energy models is the analysis of the $ \omega_D - \omega'_D $ plane, where $\omega'_{D}=d\omega_{D}/d~ln~a$ represents evolution of the EoS parameter with respect to the logarithm of the scale factor. $ \omega_D - \omega'_D $ analysis has become a standard tool for testing theoretical dark energy scenarios, including quintessence, phantom, quintom, holographic dark energy, agegraphic dark energy, and various modified gravity models. This diagnostic was introduced by Caldwell and Linder \cite{lindd} as a model-independent framework to characterize the dynamical behavior of dark energy. In this phase-space analysis, different dark energy scenarios occupy distinct regions, allowing a clear comparison of their evolutionary trajectories. A cosmological constant corresponds to the fixed point $(\omega_D - \omega'_D)=(-1,0)$ in the plane, whereas dynamical dark energy models generally evolve away from or toward this point. This method was first applied to the quintessence DE model, which divides the plane into two distinct regions: thawing models, which evolve away from the cosmological constant state $(\omega_D - \omega'_D)=(-1,0)$, and freezing models, which gradually approach it during cosmic evolution. It has been noted that the universe's expansion accelerates more rapidly in the freezing region. Consequently, the $ \omega_D - \omega'_D $ plane provides a valuable geometrical diagnostic for investigating the nature of dark energy and assessing the viability of theoretical models against observational data. Below we perform this analysis for our hybrid DE model for both the interacting and non-interacting scenarios.

The expression of $\omega'_D$ for non-interacting case is derived as
\begin{eqnarray*}
\omega'_D= \frac{3}{\left(\alpha n^2 (m-1)^2(1+z)^{-2}+c^2m^2\beta \right)} \Big{[}a_0^{-\frac{2}{m}} H_0^2 \Omega_{m0}(3-\frac{2}{m}) (1+z)^{4-\frac{2}{m}}
\end{eqnarray*}
\begin{equation}\label{dashedomegaNONINT}
\frac{2\alpha (m-1)^2n^2  }{M_P^2 (1+z)(\alpha n^2 (m-1)^2(1+z)^{-2}+c^2m^2\beta)} \Big{(} 3 m (m-1)+a_0^{-\frac{2}{m}} H_0^2 M_P^2 \Omega_{m0} (1+z)^{3-\frac{2}{m}}\Big{)} \Big{]}
\end{equation}

Similarly, the expression of $\omega'_D$ for the interacting case is obtained as,
\begin{eqnarray*}
\omega'_D= \frac{a_0^{-\frac{2}{m}}}{3\left(\alpha n^2 (m-1)^2(1+z)^{-2}+c^2m^2\beta \right)} \Big{[} H_0^2 \Omega_{m0}(3-\frac{2}{m}-3\delta) (1+z)^{4-3\delta-\frac{2}{m}}
\end{eqnarray*}
\begin{equation}\label{dashedomegaINT}
\frac{2\alpha (m-1)^2n^2  }{M_P^2 (1+z)(\alpha n^2 (m-1)^2(1+z)^{-2}+c^2m^2\beta)} \Big{(} 2 m (m-1)a_0^{\frac{2}{m}}+ H_0^2 M_P^2 \Omega_{m0} (1+z)^{3-3\delta-\frac{2}{m}}\Big{)} \Big{]}
\end{equation}

To plot the $( \omega_D - \omega'_D )$ plane for non-interacting and interacting cases, we employ equations (\ref{omegaNONINT}), (\ref{dashedomegaNONINT}),(\ref{omegaINT}), and (\ref{dashedomegaINT}).

\begin{figure}[hbt!]
\begin{center}
\includegraphics[height=4in,width=2.8in]{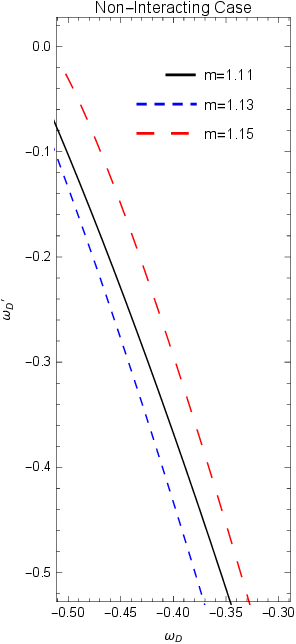}\\
\vspace{3mm}
\caption{The figure shows the plot of $\omega_D- \omega'_D $ for the hybrid model for the non-interacting case. Other parameters are considered as $H_0=69$, $\Omega_{m0}=0.29$, $a_0=1.2925$, $\alpha=0.94$, $\beta=0.12$ and $n=2.5$.}
\label{figscale41}
\end{center}
\end{figure}

\begin{figure}[hbt!]
\begin{center}
\includegraphics[height=4in,width=3.2in]{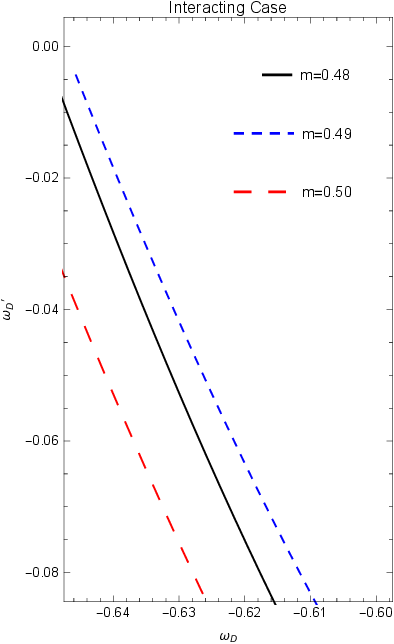}\\
\vspace{3mm}
\caption{The figure shows the plot of $\omega_D- \omega'_D $ for the hybrid model for the interacting case. Other parameters are considered as $H_0=69$, $\Omega_{m0}=0.29$, $a_0=1.2925$, $\alpha=0.94$, $\beta=0.12$ and $n=2.5$.
}
\label{figscale42}
\end{center}
\end{figure}

The plot of $\omega_D- \omega'_D$ for the non-interacting and interacting cases is presented in figures (\ref{figscale41}) and (\ref{figscale42}) respectively for different values of the power-law parameter $m$. We observe that for all values of $m$ the trajectories for both models exhibit a freezing behavior since $\omega'_{D}<0$ and $\omega'_{D}\to 0$ as $\omega_{D}\to -1$. For both scenarios, the trajectories evolve toward the cosmological constant state with $\omega'_{D}<0$. This behavior implies that the DE component was more dynamic in the past, but eventually became more and more similar to a cosmological constant. From an observational standpoint, the model's freezing nature is especially interesting because dark energy theories with equation of state parameters that stay near (-1) are preferred by current cosmological data. As a result, the evolutionary trajectory's location in the freezing region validates both the model's stable late-time behavior and shows that it is consistent with the universe's late accelerated expansion. The model is a good contender to describe the dark energy sector, since it naturally converges to a cosmological-constant-like state in the future, as indicated by the approach toward the $\Lambda$CDM fixed point.

\subsection{Statefinder parameters: $r-s$ plane}
Over time, numerous dark energy models have been introduced to account for the accelerated expansion of the universe, giving rise to the challenge of distinguishing them from one another. To assess the viability of these models, the statefinder parameters are commonly employed \cite{Sahni:2002fz, Alam:2003sc}. These parameters are dimensionless, and the corresponding cosmological plane is known as the $r-s$ plane. The trajectories in this plane indicate how far a given DE model deviates from the $\Lambda$CDM limit. The key regions defined by these parameters are: $(r, s) = (1, 0)$ representing the $\Lambda$CDM limit, $(r, s) = (1, 1)$ denoting the cold dark matter limit, $s > 0, r < 1$ corresponding to the quintessence region, and $s < 0, r > 1$ indicating the phantom region.

The statefinder parameters are expressed as
\begin{equation}
r = \frac{\dddot{a}}{aH^3}, \quad s = \frac{r - 1}{3(q - \tfrac{1}{2})}, 
\end{equation}
where $q$ represents the deceleration parameter. In terms of the Hubble and deceleration parameters, the statefinder parameters can be rewritten as
\begin{equation}
r = 2q^2 + q - \frac{\dot{q}}{H},~~~ \quad s = \frac{r - 1}{3(q - \tfrac{1}{2})}.  
\end{equation}

\subsubsection{Non-interacting Case}
For the hybrid DE model, the expressions for the statefinder parameters for the non-interacting scenario are given by
\begin{equation}
r= -1 + \frac{M_P^2}{m^3}\left[ 2 \alpha n^2 (m-1)^3(1+z)^{-2}+A_1\right]+2\left[1+ \frac{M_P^2}{m^3}A_1\right]^2
\end{equation}
and 
\begin{equation}
s=- \frac{2 M_P^2}{3}\frac{\left( 2 \alpha n^2 (m-1)^3 (1+z)^{-2}+5 A_1+2\frac{M_P^2 }{m^3}A_1^2\right)}{3 m^3+2 M_P^2 A_1},
\end{equation}
where $A_1=\alpha n^2 (m-1)^3 (1+z)^{-2}-c^2 m^2 \beta$

\begin{figure}[hbt!]
\begin{center}
\includegraphics[height=2in,width=3.1in]{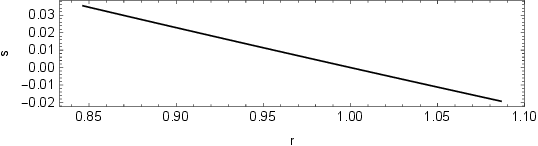}\\
\vspace{3mm}
\caption{The figure shows the plot of $r-s$ for the hybrid model for the non-interacting case. Other parameters are considered as $H_0=69$, $\Omega_{m0}=0.29$, $a_0=1.273$, $\alpha=0.74$, $\beta=0.12$, $n=2.5$ and $m=1.148$.}
\label{figscale31}
\end{center}
\end{figure}

The plot of $r-s$ for the non-interacting case is shown in figure (\ref{figscale31}). We see that for $r=1$ we have $s=0$, showing the $\Lambda$CDM point. We see that the $r-s$ plane can be divided into two different parts around this standard $\Lambda$CDM point. For $r>1$, we see that $s<0$, showing a phantom-like evolution. In the region $r<1$ we have $s>0$, showing a quintessence-like evolution. So our hybrid model exhibits both type of evolution in the $r-s$ plane. But if we consider the whole plane we see that the region $r<1$, $s>0$ dominates over the region $r>1$, $s<0$. This shows the dominance of the quintessence regime for the hybrid DE model.

\subsubsection{Interacting Case}
For the hybrid DE model, the expressions for the statefinder parameters for the interacting scenario are given by
\begin{eqnarray*}
r=\Big{[}(9 (\delta -1) \delta +2) m^2 \rho_{m0} (z+1)^{-3 \delta }+6 a_0^3 (m-1) M_P^2 \left(a_0^{-1/m} (z+1)^{-1/m}\right)^{3 m-2} \Big{(}\alpha  a_0^2 (m-1) 
\end{eqnarray*} 
\begin{eqnarray*}
(2 m-1) (3 m-2) n^2 a_0^{-2} (z+1)^{-2}+\beta  c^2 (m-2) m^2\Big{)}\Big{]}\Big{[}2 m^2 \Big(\rho_{m0} (z+1)^{-3 \delta}
\end{eqnarray*}
\begin{equation}    
+3 a_0^3 M_P^2 \left(a_0^{-1/m} (z+1)^{-1/m}\right)^{3 m-2} \left(\alpha (m-1)^2 n^2 (z+1)^{-2}+\beta  c^2 m^2\right)\Big{]}^{-1}
\end{equation}

\begin{eqnarray*}
s=-\Big[3 (\delta -1) \delta  m^2 \rho_{m0}  (z+1)^{-3 \delta }+2 a_0^3 M_P^2 \left(a_0^{-1/m} (z+1)^{-1/m}\right)^{3 m-2} \Big(\alpha  n^2(m-1)^3 (5 m-2)  (z+1)^{-2}
\end{eqnarray*}
\begin{eqnarray*}
+\beta  c^2 m^2 (2-3 m)\Big)\Big]\times   \Big[3 m \Big(\delta  m \rho_{m0}  (z+1)^{-3 \delta }+a_0^3 M_P^2 \left(a_0^{-1/m} (z+1)^{-1/m}\right)^{3 m-2}
\end{eqnarray*}
\begin{equation}
\left(\alpha n^2 (m-1)^2 (5 m-2) (z+1)^{-2}+\beta  c^2 m^2 (3 m-2)\right)\Big)\Big]^{-1}
\end{equation}
    
\begin{figure}[hbt!]
\begin{center}
\includegraphics[height=3.5in,width=4in]{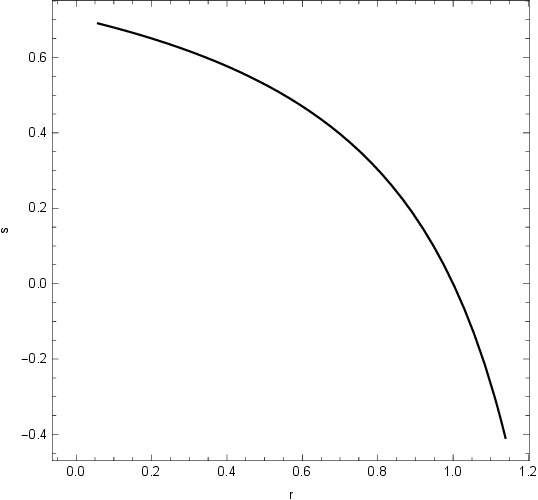}\\
\vspace{3mm}
\caption{The figure shows the plot of $r-s$ for the hybrid model for the interacting case. Other parameters are considered as $H_0=69$, $\Omega_{m0}=0.29$, $a_0=1.502$, $\alpha=0.991$, $\beta=0.0508$, $n=2.5$ and $m=0.49$.}
\label{figscale32}
\end{center}
\end{figure}

The plot of $r-s$ for the interacting case is shown in Figure (\ref{figscale32}). We see that for $r=1$ we have $s=0$, showing the $\Lambda$CDM point. We see that just like the non-interacting case, here also the $r-s$ plane can be divided into two different parts around this standard $\Lambda$CDM point. One part shows a quintessence-like evolution and the other part shows a phantom-like evolution.

\subsection{Stability Analysis: Squared Speed of Sound}
For the hybrid dark energy model under consideration, we further analyze its classical stability against small perturbations through the squared speed of sound.
The expression for the squared speed of sound reads:
\begin{equation}\label{vs1}
v_s^2=\frac{\dot{p}_D}{\dot{\rho}_D}=\frac{\acute{p_D}}{\acute{\rho_D}}
\end{equation}
where 'dot' represents the time derivative, and 'dash' represents the derivative with respect to the scale factor $a$. The expression (\ref{vs1}) can be rewritten as:
\begin{equation}\label{vs2}
v_s^2=\omega_D+\omega'_D \frac{\rho_D}{\rho'_D}
\end{equation}
The sign of $v_{s}^{2}$ determines the stability of the model: a positive value ($v_{s}^{2}>0$) corresponds to classical stability under linear perturbations, whereas a negative value ($v_{s}^{2}<0$) signals classical instability.
\begin{figure}[hbt!]
\begin{center}
\includegraphics[height=2.5in]{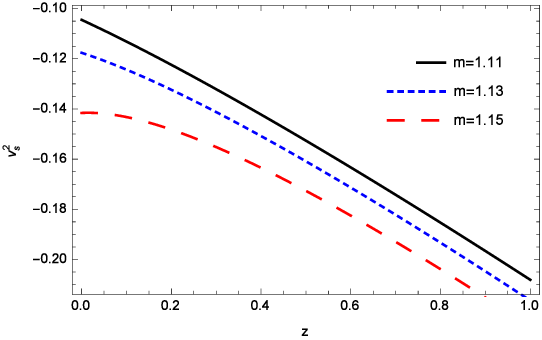}\\
\vspace{3mm}
\caption{The figure shows the plot of $v_s^2$ for the hybrid model for the non-interacting case. Other parameters are considered as $H_0=69$, $\Omega_{m0}=0.29$, $a_0=1.273$, $\alpha=1.009$, $\beta=0.07$ and $n=2.5$.}
\label{figscale61}
\end{center}
\end{figure}
\begin{figure}[hbt!]
\begin{center}
\includegraphics[height=2.5in]{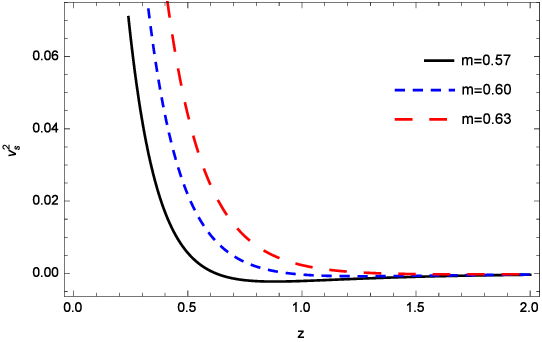}\\
\vspace{3mm}
\caption{The figure shows the plot of $v_s^2$ for the hybrid model for the interacting case. Other parameters are considered as $H_0=69$, $\Omega_{m0}=0.29$, $a_0=1.502$, $\alpha=0.991$, $\beta=0.0508$ and $n=2.5$.}
\label{figscale62}
\end{center}
\end{figure}

The expression of $v_s^2$ for the non-interacting and interacting cases can be obtained using equations (\ref{omegaNONINT}), (\ref{dashedomegaNONINT}) and (\ref{DE Expression}) ; (\ref{omegaINT}), (\ref{dashedomegaINT}), (\ref{DE Expression}) respectively. The plot of $v_s^2$ for the non-interacting and interacting cases is shown in figures (\ref{figscale61}) and (\ref{figscale62}) respectively. We observe that for the non-interacting case the model is classically unstable ($v_{s}^{2}<0$) and the model is classically stable ($v_{s}^{2}>0$) for the interacting case. This once again iterates the importance of the interacting DE models over the non-interacting counterparts.

\section{Constraints on model parameters: Observational data analysis}
In this section we perform observational data analysis using current cosmological data to constrain the parameter space of the hydrid DE model. In order to do this analysis we consider different data sets like the Hubble data from Cosmic Chronometers (CC) \cite{Jimenez2002, Moresco2015}, BAO data, DESI data. The covariance matrix used to estimate the likelihood is constructed following the approach described in \cite{Moresco2012, Moresco2016}. These data points provide direct constraints on the expansion history of the universe. Below we discuss the different datasets that will be used to perform the analysis.

\begin{itemize}
   \item \textbf{HUBBLE Data:} We use 32 model-independent measurements of the Hubble parameter
    $H(z)$, commonly known as Cosmic Chronometers (CC) \cite{Jimenez2002,Moresco2015}. The covariance matrix used to estimate the likelihood is constructed following the approach described in \cite{Moresco2012,Moresco2016}. These data points provide direct constraints on the expansion history of the universe.

    \item \textbf{BAO Data:} The BAO dataset provides constraints on the large-scale structure of the universe through measurements of the baryon acoustic feature imprinted in the matter power spectrum. We use the distance measurements from various galaxy surveys compiled in the DESI and eBOSS analyses \cite{DESI2024, eBOSS2020}. The observables considered include \( D_M/r_d \), \( D_H/r_d \), and \( D_V/r_d \), where \( D_M \) is the comoving angular diameter distance, \( D_H \) the Hubble distance, and \( r_d \) the comoving sound horizon at the drag epoch.
    \item \textbf{DESI Data:} The Dark Energy Spectroscopic Instrument (DESI) Release II dataset provides high-precision BAO and redshift-space distortion (RSD) measurements across a wide redshift range \cite{DESI2025}. These measurements significantly improve constraints on late-time cosmic acceleration and expansion rate, complementing the HUBBLE and BAO datasets.
\end{itemize}

\subsection{Statistical Analysis}
The model parameters are constrained using a Markov Chain Monte Carlo (MCMC) approach implemented via the \texttt{emcee} sampler \cite{ForemanMackey2013}. The posterior distributions are visualized using the \texttt{GetDist} package \cite{Lewis2019}. For each dataset and their combinations, we compute the best-fit values and corresponding \( 1\sigma \) confidence intervals (68\% C.L.), summarized in Tables 1 and 2 for the non-interacting and the interacting cases respectively.

The likelihood function is defined as
\begin{equation}
    \mathcal{L} \propto \exp\left(-\frac{1}{2}\chi^2\right)
\end{equation}
where the chi-square function is given by
\begin{equation}
    \chi^2 = \Delta \mathbf{D}^T \mathbf{C}^{-1} \Delta \mathbf{D}
\end{equation}
with \(\Delta \mathbf{D} = \mathbf{D}_{\text{obs}} - \mathbf{D}_{\text{th}}\), and \(\mathbf{C}\) denoting the covariance matrix for each dataset.

\subsection{Contour Representation and Colour Scheme}
\begin{figure}[h!]
    \centering
    \includegraphics[width=\linewidth]{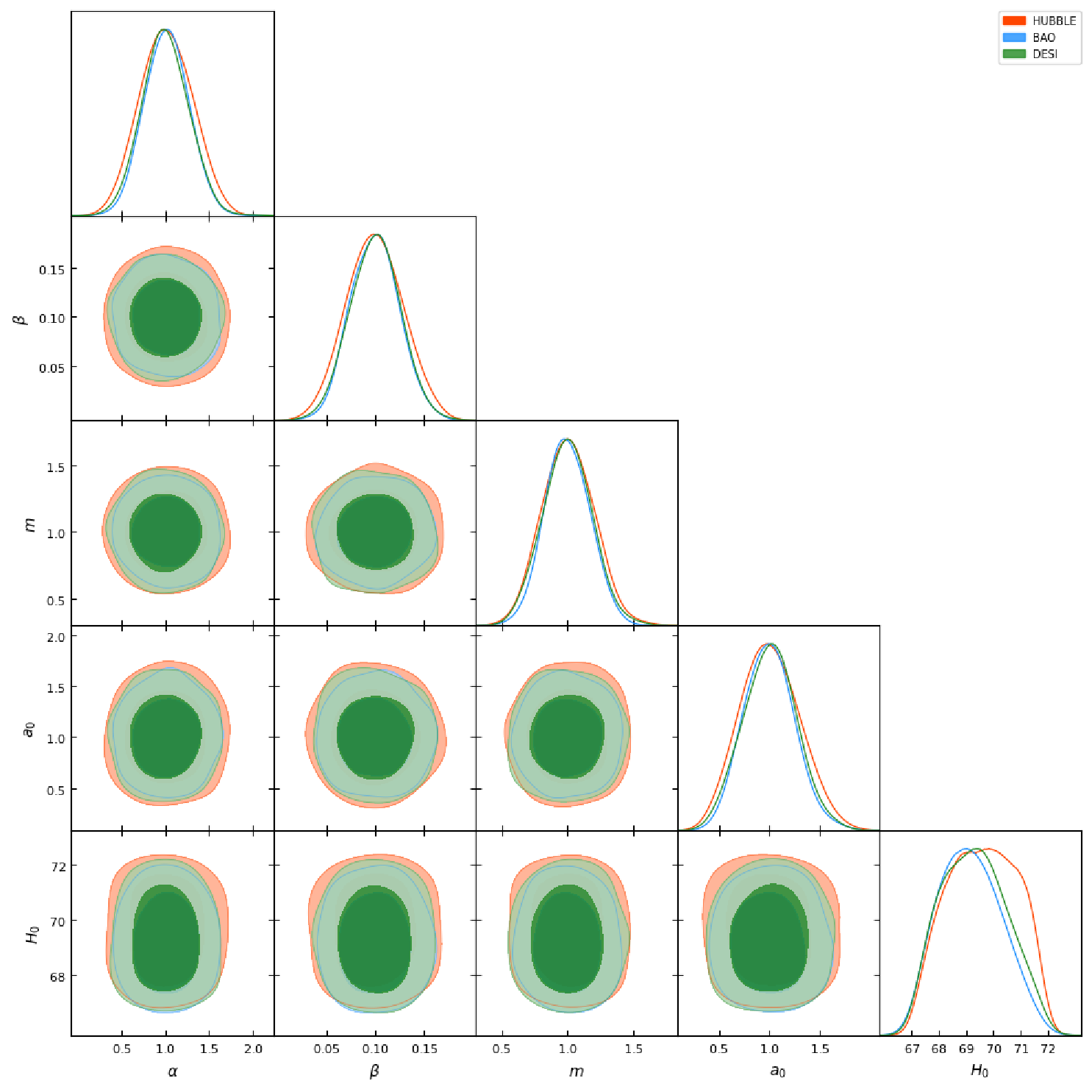}
    \caption{Joint and marginalized posterior distributions  with combinations of datasets Hubble, BAO and DESI for Non-interacting Case with fixed parameters $M_{P} = 1$, $\Omega_{m0} = 0.29$, $n = 1.1$ and $c = 0.76$.}
    \label{gg1}
\end{figure}

\begin{figure}[h!]
    \centering
    \includegraphics[width=\linewidth]{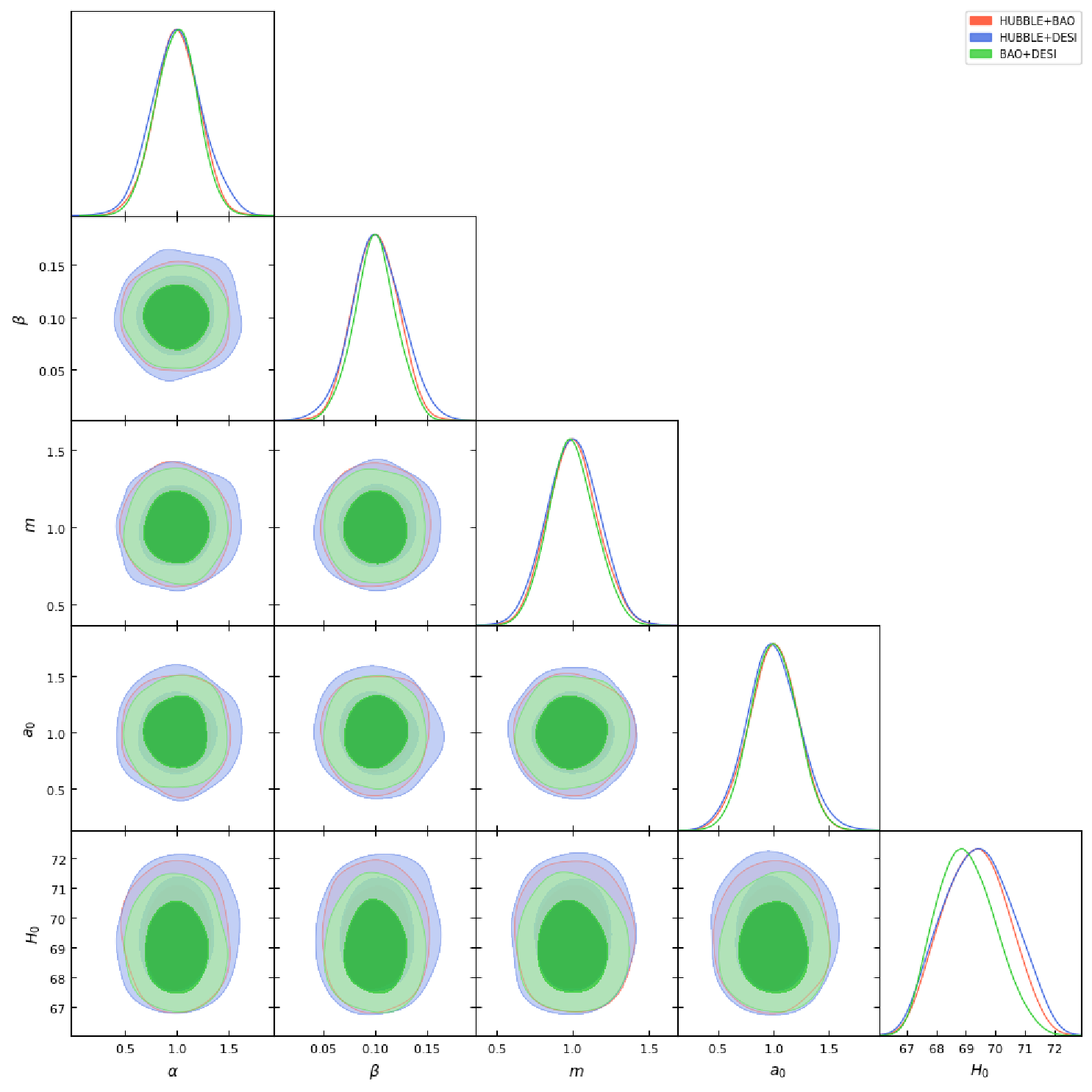}
    \caption{Joint and marginalized posterior distributions  with combinations of  datasets combination of  Hubble+BAO, Hubble+DESI, BAO+DESI for Non-interacting Case with fixed parameters $M_{P} = 1$, $\Omega_{m0} = 0.29$, $n = 1.1$ and $c = 0.76$.}
    \label{gg2}
\end{figure}

\begin{figure}[h!]
    \centering
    \includegraphics[width=\linewidth]{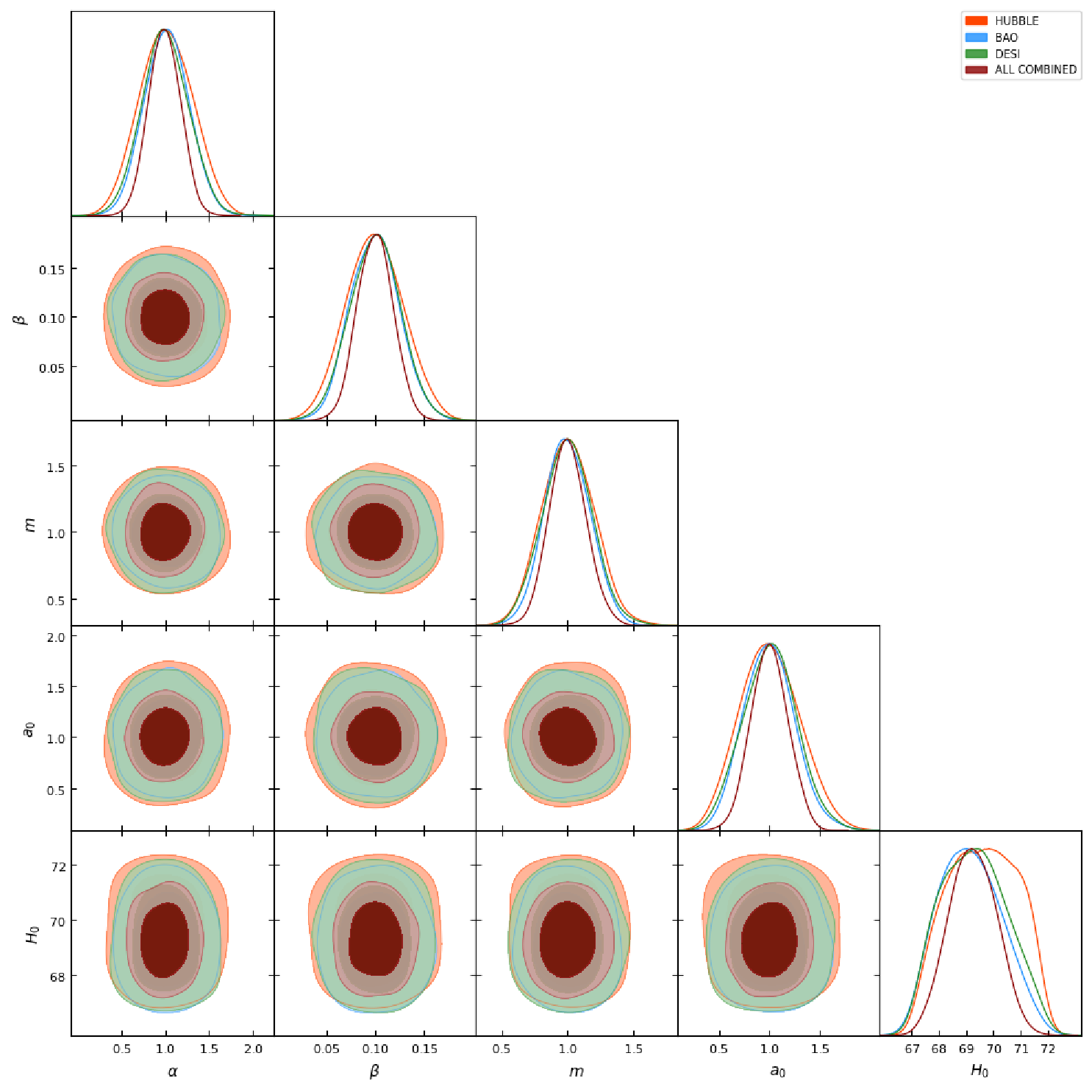}
    \caption{Joint and marginalized posterior distributions  with combinations of  datasets Hubble, BAO, DESI and Combined (Hubble+BAO+DESI) for Non-interacting Case  with fixed parameters $M_{P} = 1$, $\Omega_{m0} = 0.29$, $n = 1.1$ and $c = 0.76$}
    \label{gg3}
\end{figure}

\begin{figure}[h!]
    \centering
    \includegraphics[width=\linewidth]{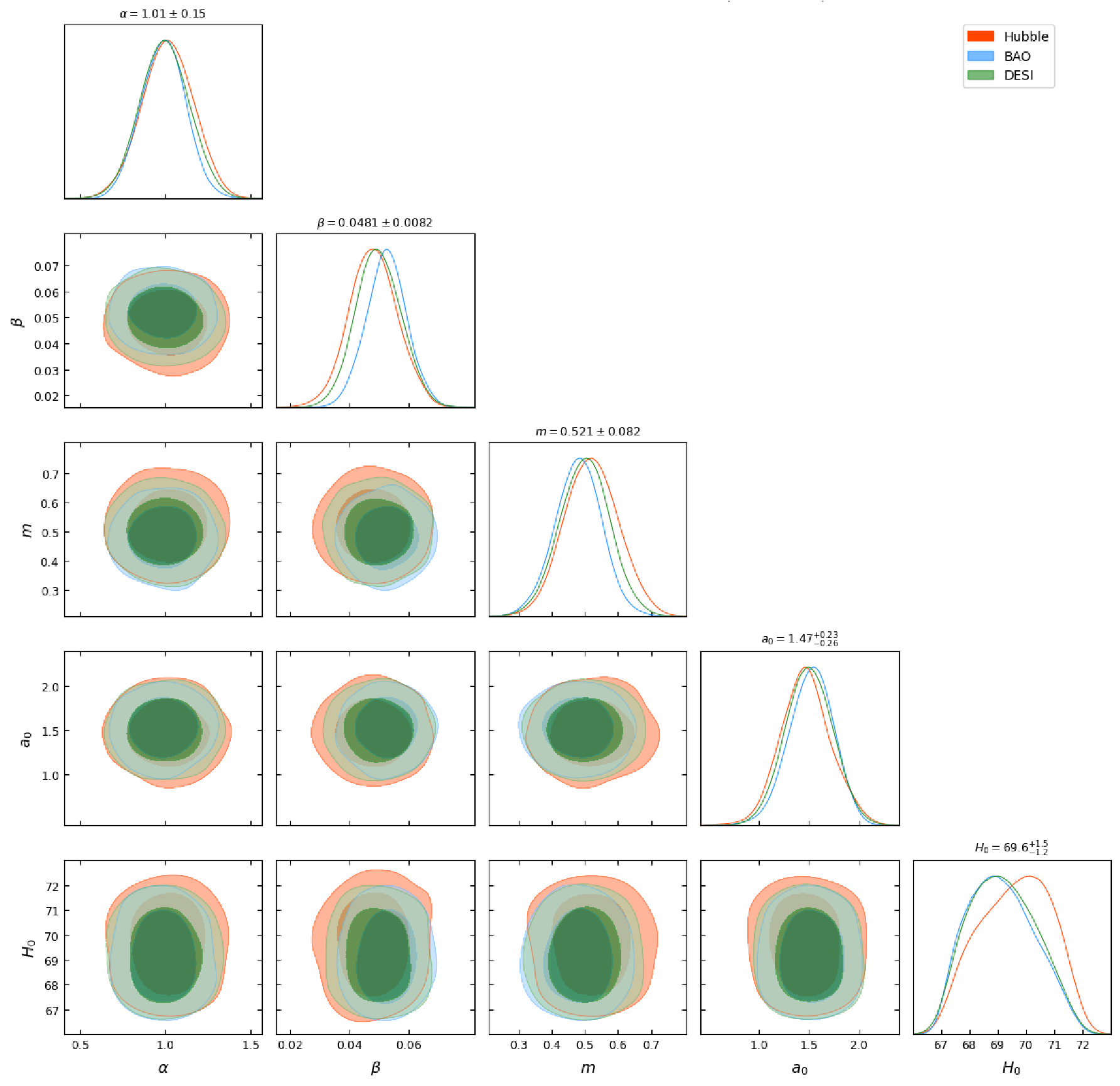}
    \caption{Joint and marginalized posterior distributions  with combinations of datasets Hubble, BAO and DESI for Interacting Case with fixed parameters taken as $M_P = 1$, $\Omega_{m0} = 0.29$, $n = 2.5$, $c = 0.76$, and $\delta = 0.02$.}
    \label{gg4}
\end{figure}

\begin{figure}[h!]
    \centering
    \includegraphics[width=\linewidth]{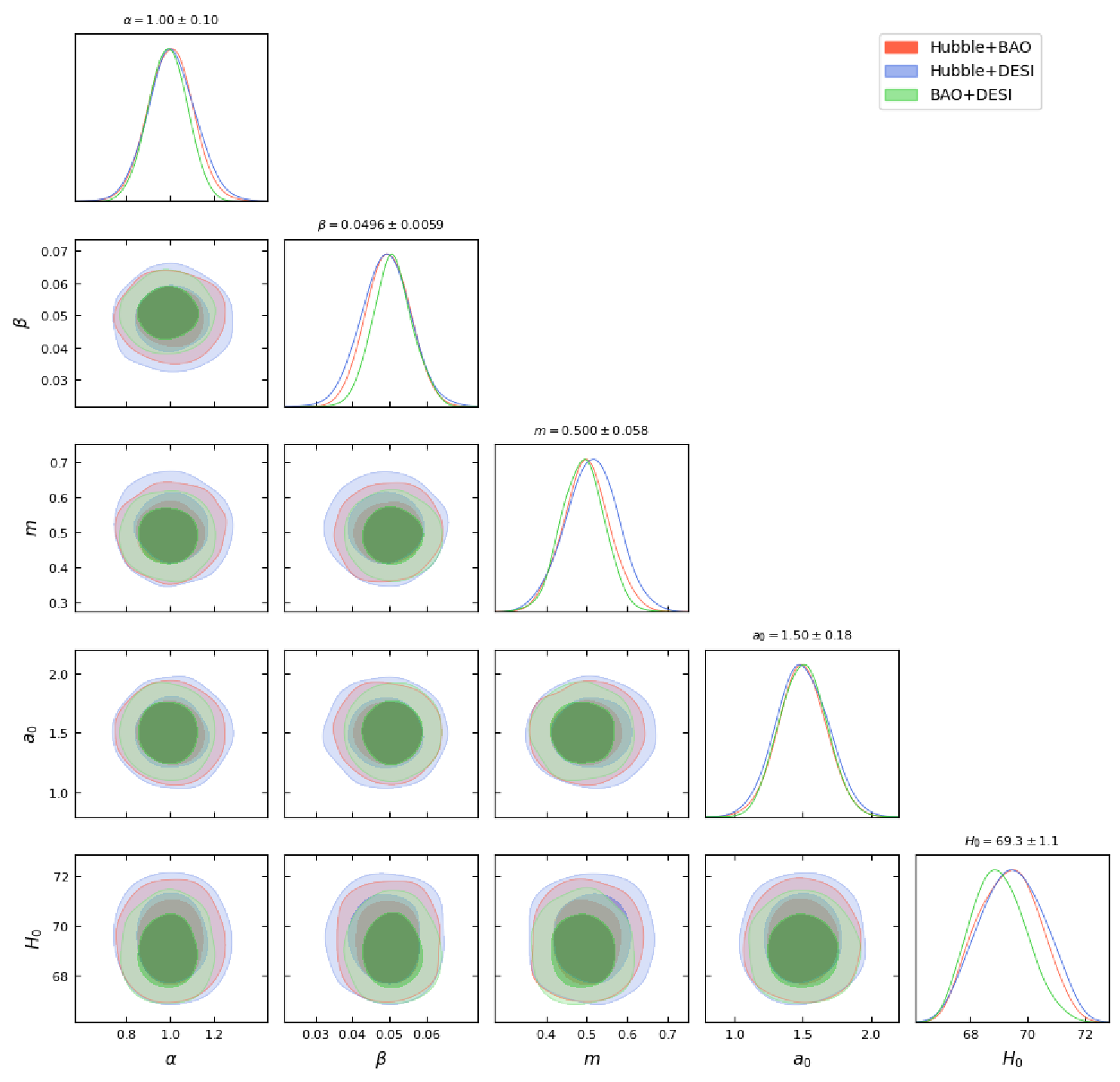}
    \caption{Joint and marginalized posterior distributions  with combinations of datasets Hubble, BAO, DESI and Combined (Hubble+BAO+DESI) for Interacting Case with fixed parameters taken as $M_P = 1.0$, $\Omega_{m0} = 0.29$, $n = 2.5$, $c = 0.76$, and $\delta = 0.02$.}
    \label{gg5}
\end{figure}

\begin{figure}[h!]
    \centering
    \includegraphics[width=\linewidth]{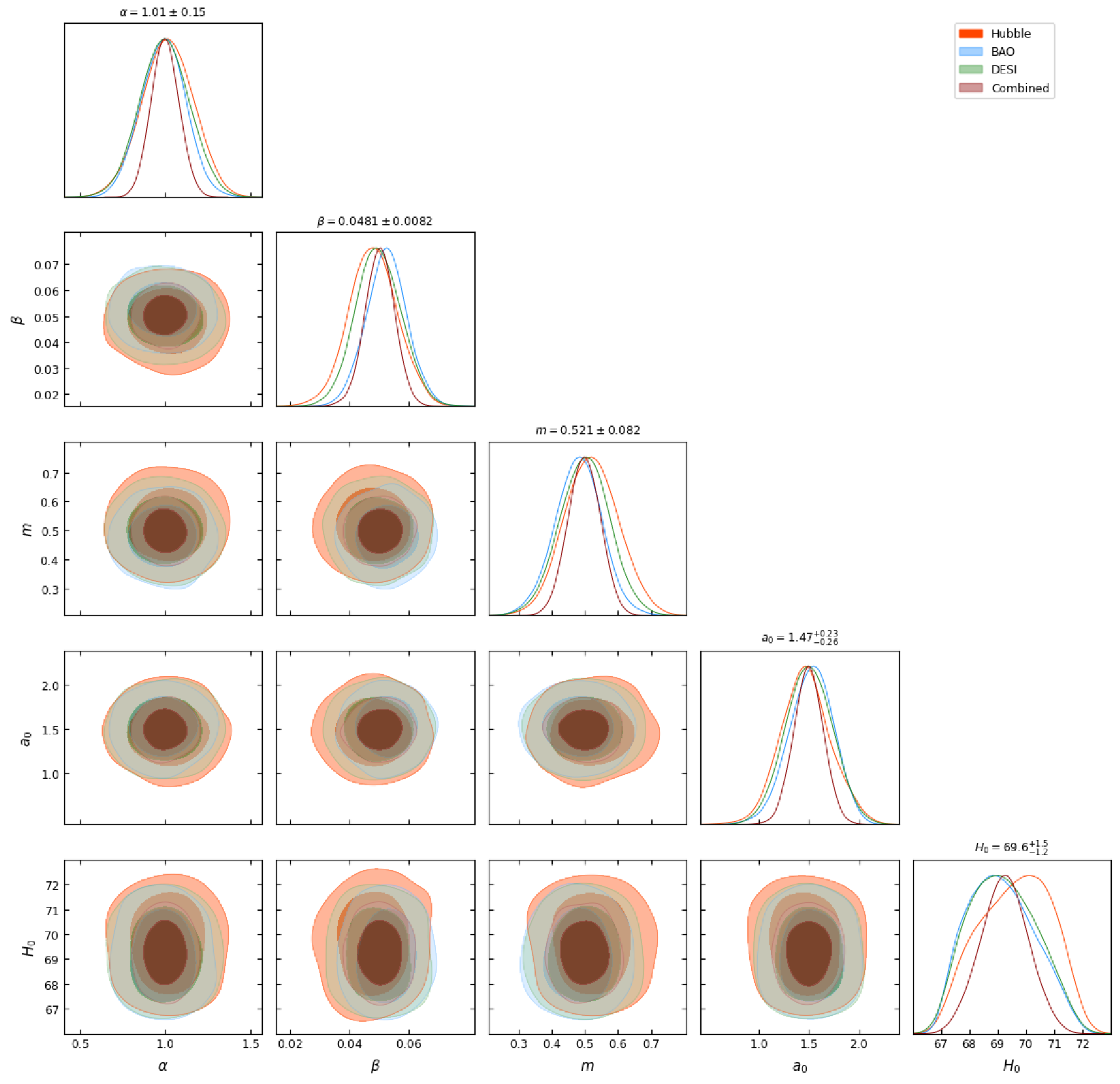}
    \caption{Joint and marginalized posterior distributions  with combinations of  datasets Hubble, BAO, DESI, and Combined (Hubble+BAO+DESI) for Interacting Case with fixed parameters taken as $M_P = 1$, $\Omega_{m0} = 0.29$, $n = 2.5$, $c = 0.76$, and $\delta = 0.02$.}
    \label{gg6}
\end{figure}

In the figures (\ref{gg1}), (\ref{gg2}), (\ref{gg3}), (\ref{gg4}), (\ref{gg5}) and (\ref{gg6}) we have generated the two-dimensional confidence contours showing the model parameter constraints from different observational datasets and their combinations. The confidence contours shown in the parameter constraint plots represent the joint probability distributions of the model parameters obtained from the likelihood analysis. The color scheme of the contour plots follows the standard cosmological convention:
\begin{itemize}
    \item The \textbf{light-shaded region} corresponds to the \textbf{95\% confidence level (C.L.)}, indicating the parameter space within which the true values are expected to lie with 95\% probability.
    \item The \textbf{dark-shaded region} corresponds to the \textbf{68\% confidence level (C.L.)}, representing the $1\sigma$ range of the best-fit values.
\end{itemize}

The central black dot in each contour marks the \textbf{best-fit parameter values}, while the surrounding elliptical regions illustrate the covariance between the two fitted parameters. The darker inner contour thus represents the most probable region of parameter space consistent with the observational data, while the lighter outer contour shows the extended range allowed by the data at a higher uncertainty level. All contour plots in this analysis ensure statistical consistency and reproducibility. The colors were chosen to maintain clarity between the 68\% and 95\% confidence regions across all cases.

\begin{table}[htbp]
\centering
\caption{Parameter Constraints Summary for the Non-interacting Case}
\resizebox{\textwidth}{!}{%
\begin{tabular}{lccccc}
\hline\hline
Dataset & $\alpha$ & $\beta$ & $m$ & $a_0$ & $H_0$ [km s$^{-1}$ Mpc$^{-1}$] \\
\hline
HUBBLE          
& $1.0121 \pm 0.3021$ & $0.0994 \pm 0.0294$ & $1.0050 \pm 0.1985$ & $1.0052 \pm 0.2980$ & $69.61 \pm 1.27$ \\

BAO             
& $1.0167 \pm 0.2548$ & $0.0995 \pm 0.0252$ & $0.9987 \pm 0.1719$ & $0.9990 \pm 0.2538$ & $69.15 \pm 1.16$ \\

DESI            
& $1.0024 \pm 0.2685$ & $0.0997 \pm 0.0261$ & $0.9987 \pm 0.1882$ & $1.0057 \pm 0.2683$ & $69.29 \pm 1.22$ \\

HUBBLE+BAO      
& $0.9990 \pm 0.2153$ & $0.1004 \pm 0.0216$ & $1.0033 \pm 0.1637$ & $0.9966 \pm 0.2181$ & $69.34 \pm 1.09$ \\

HUBBLE+DESI     
& $1.0030 \pm 0.2465$ & $0.1019 \pm 0.0247$ & $1.0032 \pm 0.1709$ & $0.9981 \pm 0.2376$ & $69.42 \pm 1.18$ \\

BAO+DESI        
& $0.9948 \pm 0.2044$ & $0.1001 \pm 0.0202$ & $0.9965 \pm 0.1530$ & $1.0041 \pm 0.2076$ & $69.01 \pm 0.99$ \\

HUBBLE+BAO+DESI 
& $0.9921 \pm 0.1831$ & $0.0998 \pm 0.0183$ & $1.0010 \pm 0.1415$ & $1.0024 \pm 0.1826$ & $69.24 \pm 0.87$ \\
\hline
\end{tabular}}
\label{tab:parameter_constraints_noninteracting}
\end{table}

\begin{table}[htbp]
\centering
\caption{Parameter Constraints Summary for the Interacting Case}
\resizebox{\textwidth}{!}{%
\begin{tabular}{lccccc}
\hline\hline
Dataset & $\alpha$ & $\beta$ & $m$ & $a_0$ & $H_0$ [km s$^{-1}$ Mpc$^{-1}$] \\
\hline
HUBBLE          
& $1.014 \pm 0.151$ & $0.0481 \pm 0.0082$ & $0.521 \pm 0.082$ & $1.471 \pm 0.251$ & $69.64 \pm 1.27$ \\

BAO             
& $0.988 \pm 0.130$ & $0.0526 \pm 0.0069$ & $0.480 \pm 0.071$ & $1.523 \pm 0.223$ & $69.13 \pm 1.18$ \\

DESI            
& $0.997 \pm 0.144$ & $0.0499 \pm 0.0077$ & $0.500 \pm 0.076$ & $1.501 \pm 0.234$ & $69.19 \pm 1.20$ \\

HUBBLE+BAO      
& $1.003 \pm 0.102$ & $0.0496 \pm 0.0059$ & $0.500 \pm 0.058$ & $1.496 \pm 0.176$ & $69.35 \pm 1.07$ \\

HUBBLE+DESI     
& $1.009 \pm 0.110$ & $0.0490 \pm 0.0067$ & $0.513 \pm 0.065$ & $1.497 \pm 0.191$ & $69.47 \pm 1.13$ \\

BAO+DESI        
& $0.991 \pm 0.089$ & $0.0508 \pm 0.0053$ & $0.490 \pm 0.053$ & $1.502 \pm 0.171$ & $68.99 \pm 0.96$ \\

HUBBLE+BAO+DESI 
& $0.999 \pm 0.082$ & $0.0503 \pm 0.0050$ & $0.498 \pm 0.049$ & $1.496 \pm 0.150$ & $69.25 \pm 0.84$ \\
\hline
\end{tabular}}
\label{tab:parameter_constraints_modified_gravity}
\end{table}
\section{Hubble Tension Analysis}
\label{sec:hubble_tension}

The Hubble tension stands as one of the most significant unresolved challenges in modern cosmology. It refers to the persistent discrepancy in the inferred value of the Hubble constant $H_0$, the present-day expansion rate of the universe when derived from fundamentally different observational probes, each interpreted within the framework of
the standard flat $\Lambda$CDM model. Two conflicting measurements lie at the heart of this tension. Early-universe determinations based on the Cosmic Microwave Background (CMB) anisotropies measured by the Planck satellite yield $H_0 = 67\ \mathrm{km\,s^{-1}\,Mpc^{-1}}$, while late-universe local distance-ladder measurements using Cepheid variables and Type~Ia Supernovae give $H_0 = 73\ \mathrm{km\,s^{-1}\,Mpc^{-1}}$~\cite{Planck2018,Verde2019,Riess2022}. The statistical significance of this disagreement has now grown to approximately $4$--$5\sigma$, well beyond the threshold for a chance fluctuation. This unresolved discrepancy may point toward the presence of unknown systematic errors in one or both measurements, or — more
intriguingly — may signal new physics beyond the standard cosmological paradigm~\cite{Planck2018,Verde2019,Riess2022}. Here we are interested to probe whether our constructed hybrid DE model can resolve or alleviate this tension or not.

To provide a precise and model-independent quantification of this disagreement, we adopt the standard Gaussian tension
estimator~\cite{Verde2019},
\begin{equation}
  \boxed{
  T \;=\;
  \frac{\left|\,H_0^{(1)} - H_0^{(2)}\,\right|}
       {\sqrt{\,\sigma_1^2 + \sigma_2^2\,}}
  }
  \label{eq:tension_estimator}
\end{equation}
where $H_0^{(1)}$ and $H_0^{(2)}$ denote the two Hubble constant measurements under comparison, and $\sigma_1$, $\sigma_2$ are their respective $1\sigma$ uncertainties. The dimensionless quantity $T$ expresses the level of disagreement in units of the combined standard deviation.

In the present work, we evaluate the Hubble tension for both physical scenarios considered: the non-interacting and interacting hybrid HDE--NADE model. In each case, the best-fit value of $H_0$ obtained from the full combined dataset HUBBLE+BAO+DESI — already reported in the tables is compared to a $\Lambda$CDM reference value derived from the same observational datasets. This ensures a fair, dataset-consistent comparison between the hybrid dark energy model and the standard cosmological framework.

\subsection{Hubble Tension from Best-Fit Values}
In this analysis we will use the best-fit parameter values that have already been found in the previous section. The reference value of the Hubble parameter from the $\Lambda$CDM model, computed using the same observational datasets as employed in our MCMC analysis, is
\begin{equation}
  H_0^{\,\mathrm{ref}} \;=\; 71.695\,493
  \ \mathrm{km\,s^{-1}\,Mpc^{-1}}
  \label{eq:H0_ref}
\end{equation}
We now compute the Hubble tension $T$ for each of the two hybrid DE models (non-interacting and interacting) cases in turn.


\subsubsection{Non-interacting case}
\begin{figure}[htbp]
  \centering
  \includegraphics[width=0.72\textwidth]{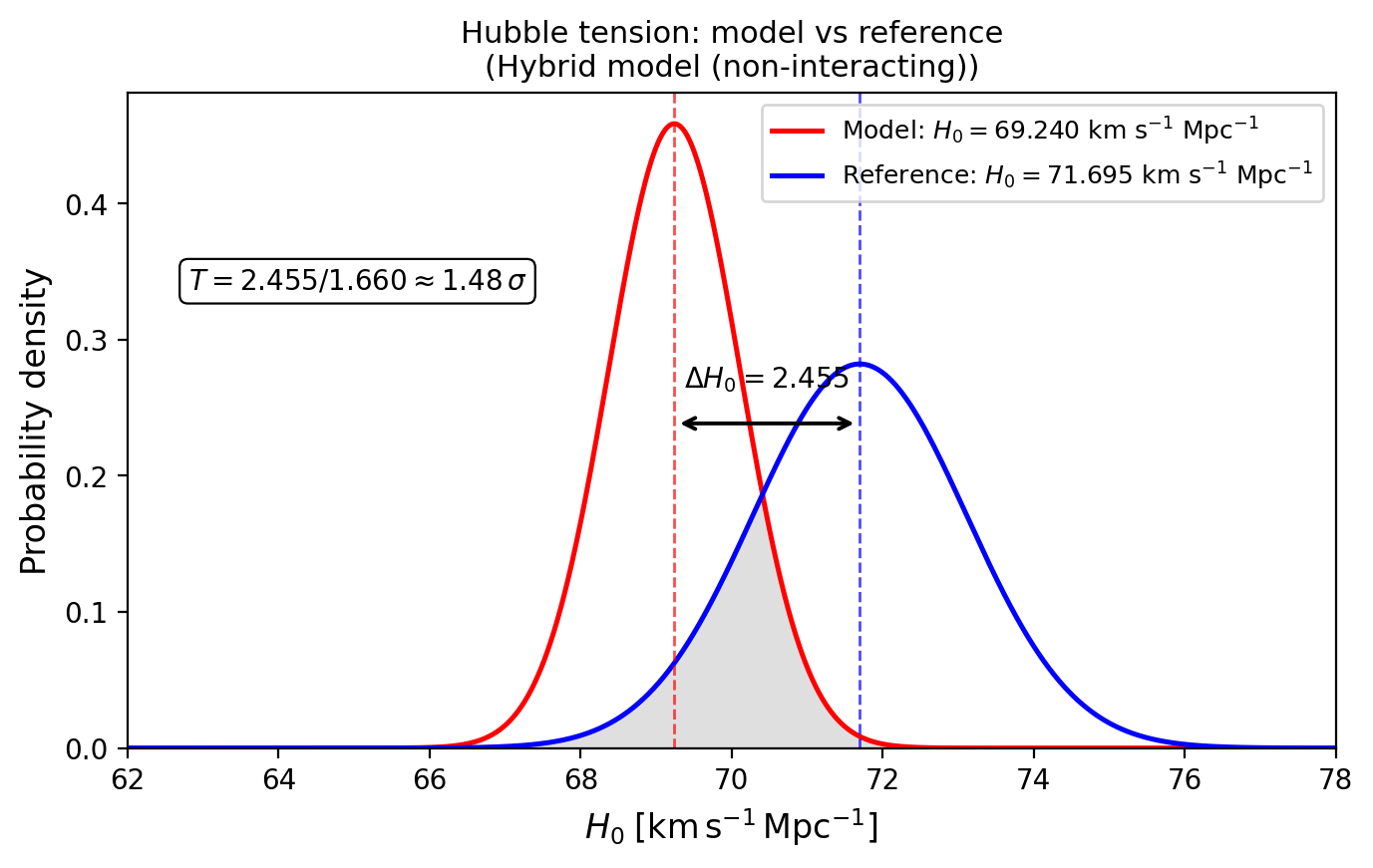}
  \caption{%
    Illustration of the Hubble tension for the non-interacting
    hybrid HDE--NADE model. The red Gaussian curve is centred on the hybrid model best-fit value of
    $H_0$.
  }
  \label{fig:tension_nonint}
\end{figure}

The best-fit Hubble parameter value obtained from the hybrid HDE--NADE model in the non-interacting scenario using the HUBBLE+BAO+DESI combined dataset is
\begin{equation}
  H_0^{\,\mathrm{model}} \;=\; 69.24
  \ \mathrm{km\,s^{-1}\,Mpc^{-1}}
  \label{eq:H0_nonint}
\end{equation}
Substituting this into tension estimator, the Hubble tension
evaluates to
\begin{equation}
  T \;\approx\;
  \frac{|\,71.695\,493 - 69.24\,|}
       {\sqrt{\,(0.87)^2 + (1.414)^2\,}}
  \;=\;
  \frac{2.455}{1.660}
  \;\approx\; \bm{1.48\,\sigma}
  \label{eq:T_nonint}
\end{equation}
The corresponding Gaussian probability distributions are shown in figure~(\ref{fig:tension_nonint}).


\subsubsection{Interacting case}
\begin{figure}[htbp]
  \centering
  \includegraphics[width=0.72\textwidth]{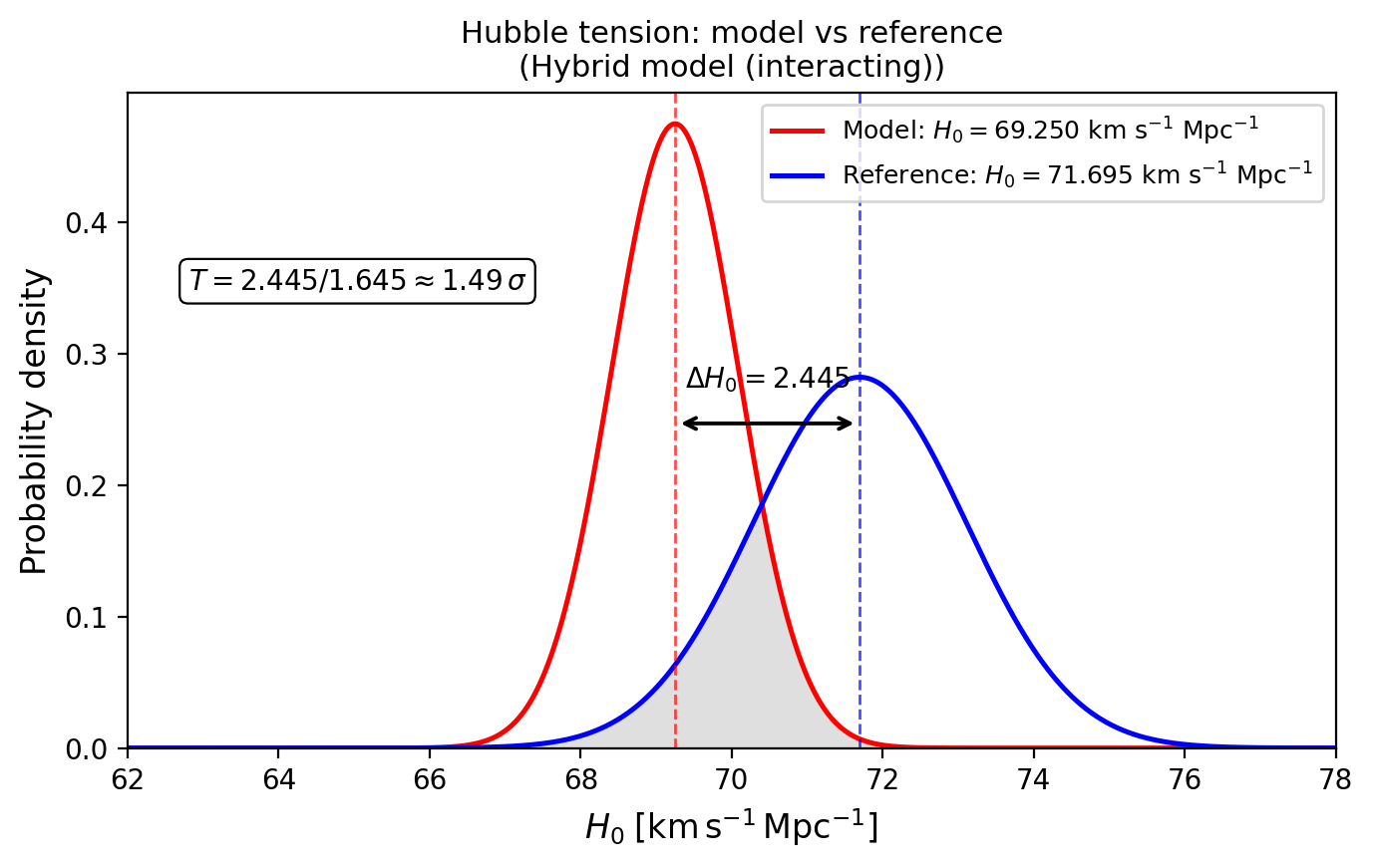}
  \caption{Illustration of the Hubble tension for the interacting hybrid HDE--NADE model. The red Gaussian curve is centred on the hybrid model best-fit value of
    $H_0$.
  }
  \label{fig:tension_int}
\end{figure}

For the interacting hybrid HDE--NADE scenario, the best-fit value of the Hubble parameter is
\begin{equation}
  H_0^{\,\mathrm{model}} \;=\; 69.25
  \ \mathrm{km\,s^{-1}\,Mpc^{-1}},
  \label{eq:H0_int}
\end{equation}
yielding a Hubble tension of
\begin{equation}
  T \;\approx\;
  \frac{|\,71.695\,493 - 69.25\,|}
       {\sqrt{\,(0.84)^2 + (1.414)^2\,}}
  \;=\;
  \frac{2.445}{1.645}
  \;\approx\; \bm{1.49\,\sigma}.
  \label{eq:T_int}
\end{equation}
This is illustrated in figure~(\ref{fig:tension_int}). 

The numerical results across both cases are remarkably consistent, with tensions clustering tightly in the narrow range $1.48\sigma$--$1.49\sigma$. This consistency across both the non-interacting and interacting dark-sector scenarios underscores the robustness of the result. While a residual discrepancy with the $\Lambda$CDM reference value persists, it is significantly milder than the $\sim 5\sigma$ tension reported between early- and late-universe
measurements in the standard framework~\cite{Planck2018,Verde2019,Riess2022}. Crucially, the tension obtained here is also substantially lower than the $\approx 2\sigma-4\sigma$ tension obtained for HDE and NADE models \cite{t1, t2}. Moreover it is lower than $\approx 2.3\sigma$ tension reported for the Ricci-Cubic Holographic Dark Energy (RCHDE) model~\cite{SanyalRudra2026}, demonstrating that the hybrid HDE--NADE construction provides a more effective partial alleviation of the Hubble tension compared to individual holographic dark energy models.

In both the figures below, the red curve represents the Gaussian probability distribution centred on the hybrid model best-fit value of $H_0$, while the blue curve corresponds to the $\Lambda$CDM reference distribution. The shaded overlap region highlights the degree of agreement between the two distributions. The double-headed arrow indicates the absolute
difference $\Delta H_0$, and the computed tension $T$ is annotated explicitly inside each panel.


\section{Discussion and Conclusion}
Given that length and time are the spatial and temporal manifestations of the same spacetime structure, it is reasonable to expect that the fundamental infrared cutoff governing dark energy should involve both quantities. This idea motivates a hybrid holographic-agegraphic paradigm, where HDE and NADE appear as complimentary components of a single spacetime-based description of dark energy. In this study, we present a hybrid dark energy model built from the combined energy densities of new agegraphic dark energy and holographic dark energy. The potential for many quantum-gravity-inspired mechanisms to simultaneously contribute to the effective dark energy sector is the driving force behind the proposed approach. Compared to the individual HDE or NADE models, this design enables us to investigate a larger class of cosmic processes.

A number of diagnostic quantities have been used to analyze the model's cosmic evolution. The behavior of the equation of state parameter suggests that the late-time accelerated expansion of the Universe can be successfully driven by the suggested model. The dark energy component may behave in a quintessence-like, phantom-like, or near cosmological-constant manner, depending on the values of the model parameters. This adaptability shows how rich the hybrid framework is and emphasizes how it can represent several expansion histories within a same scenario. In line with observational data, the deceleration parameter verifies the existence of a transition from an early decelerated epoch to the currently accelerated phase. One of the fundamental prerequisites for any workable dark energy model is such a transition. According to the obtained evolution, the hybrid contribution preserves the conventional cosmic sequence of evolution by being subdominant during the matter-dominated era and only becoming dominant at late times. We used Om diagnostic and the statefinder diagnostic analysis to further describe the model. While the paths in the statefinder plane approach the standard $\Lambda$CDM scenario given appropriate parameter selections, they clearly deviate from it during cosmic evolution for any arbitrary selection of parameter space. This shows that the hybrid model can be distinguished from traditional DE candidates using the statefinder parameters. Similarly, the Om diagnostic is a useful tool for detecting deviations from the cosmological constant paradigm and offers an independent geometrical probe of the model.

The model's stability is a key component of the current investigation. One measure of classical perturbative stability is the squared speed of sound. The obtained behavior demonstrates how the model's stability is sensitive to cosmic evolution and model parameters. The model is free of classical instabilities and can be considered physically feasible in the areas where the squared speed of sound stays positive. As a result, the stability analysis places further limitations on the permitted parameter space in addition to those obtained from background cosmological measurements. In our analysis it was found that the hybrid model is classically stable for the interacting case, whereas for the non-interacting case the model exhibits instability. This once again iterates the importance of interacting DE models over the non-interacting counterparts. We conducted a thorough statistical analysis employing cosmic chronometer, BAO, and DESI datasets to challenge the theoretical framework with observations. The observational restrictions yield best-fit estimates for the model parameters and drastically narrow the allowable parameter space. The generated confidence regions show that the suggested hybrid scenario can replicate the observed expansion history and is consistent with current cosmological data. Due to their unparalleled precision, DESI measurements are especially crucial since they enable a rigorous examination of deviations from the $\Lambda$CDM model. The hybrid model was tested against the Hubble tension issue which is a significant unresolved challenge of modern cosmology. It was found that our hybrid model do not solve the problem, but it significantly alleviates the issue, as expected from its fundamental set-up.

Overall, the findings show that a coherent framework for explaining the universe's evolution including the cosmic acceleration is provided by the hybrid HDE-NADE model. The model demonstrates rich cosmic dynamics while being consistent with existing empirical limitations by combining contributions from both holographic and agegraphic dark energy. These characteristics imply that hybrid dark energy models influenced by quantum gravity merit more research as viable substitutes for traditional dark energy scenarios. The present work is novel in that it shows that a dark energy sector made up of simultaneous holographic and agegraphic contributions can successfully accommodate recent cosmic chronometer, BAO, and DESI observations, produce a viable cosmological evolution, and remain stable within observationally allowed regions of parameter space. This implies that dark energy models influenced by hybrid quantum gravity represent an intriguing avenue for further research beyond traditional single-component dark energy scenarios. Possible future projects in this direction can be a dynamical system analysis of the model to explore the phase space. A perturbation analysis is also important to understand the behavior/stability of the model against growth perturbations.

\section*{Acknowledgments}
PR acknowledges the Inter-University Centre for Astronomy and Astrophysics (IUCAA), Pune, India for granting a visiting associateship. SMR and AS also acknowledge the hospitality provided by IUCAA during a visit, when the majority of this work was done. 

\section*{Data Availability Statement}

No data was generated or analyzed in this study.

\section*{Conflict of Interest}

There are no conflicts of interest.

\section*{Funding Statement}

There is no funding to report for this article.


\begin{thebibliography}{99}

\bibitem{perl1}
S. Perlmutter et al., \textit{Discovery of a supernova explosion at half the age of the Universe and its cosmological implications}, Nature, 391, 51--54, (1998).

\bibitem{perl2}
S. Perlmutter et al., \textit{Measurements of $\Omega$ and $\Lambda$ from 42 high redshift supernovae}, Astrophys. J., 517, 565--586, (1999).

\bibitem{perl3}
A. G. Riess et al., \textit{Observational evidence from supernovae for an accelerating universe and a cosmological constant}, Astron. J., 116, 1009--1038, (1998).

\bibitem{bao1}
D. J. Eisenstein and W. Hu, \textit{Baryonic features in the matter transfer function}, Astrophys. J., 496, 605, (1998).

\bibitem{bao2}
D. J. Eisenstein et al., \textit{Detection of the Baryon Acoustic Peak in the Large-Scale Correlation Function of SDSS Luminous Red Galaxies}, Astrophys. J., 633, 560--574, (2005).

\bibitem{cmb1}
A. A. Penzias and R. W. Wilson, \textit{A Measurement of Excess Antenna Temperature at 4080 Mc/s}, Astrophys. J., 142, 419--421 (1965).

\bibitem{cmb2}
R. H. Dicke, P. J. E. Peebles, P. G. Roll and D. T. Wilkinson, \textit{Cosmic Black-Body Radiation}, Astrophys. J., 142, 414--419 (1965).

\bibitem{brax}
P. Brax, \textit{What makes the Universe accelerate? A review on what dark energy could be and how to test it}, Rept. Prog. Phys. 81 (1) 016902 (2018).

\bibitem{Peebles2003}
P. J. E. Peebles and B. Ratra, \textit{The Cosmological Constant and Dark Energy}, Rev. Mod. Phys., 75, 559--606, (2003).

\bibitem{Padmanabhan2003}
T. Padmanabhan, \textit{Cosmological Constant: The Weight of the Vacuum}, Phys. Rep., 380, 235--320, (2003).

\bibitem{Copeland2006}
E. J. Copeland, M. Sami and S. Tsujikawa, \textit{Dynamics of Dark Energy}, Int. J. Mod. Phys. D, 15, 1753--1936, (2006).

\bibitem{Frieman2008}
J. Frieman, M. Turner and D. Huterer, \textit{Dark Energy and the Accelerating Universe}, Ann. Rev. Astron. Astrophys., 46, 385--432, (2008).

\bibitem{Caldwell2009}
R. R. Caldwell and M. Kamionkowski, \textit{The Physics of Cosmic Acceleration}, Ann. Rev. Nucl. Part. Sci., 59, 397--429, (2009).

\bibitem{Silvestri2009}
A. Silvestri and M. Trodden, \textit{Approaches to Understanding Cosmic Acceleration}, Rept. Prog. Phys., 72, 096901, (2009).

\bibitem{Li2011}
M. Li, X.-D. Li, S. Wang and Y. Wang, \textit{Dark Energy}, Commun. Theor. Phys., 56, 525--604, (2011).

\bibitem{Bamba2012}
K. Bamba, S. Capozziello, S. Nojiri and S. D. Odintsov, \textit{Dark Energy Cosmology: The Equivalent Description via Different Theoretical Models and Cosmography Tests}, Astrophys. Space Sci., 342, 155--228, (2012).

\bibitem{Li2013}
M. Li, X.-D. Li, S. Wang and Y. Wang, \textit{Dark Energy: A Brief Review}, Front. Phys., 8, 828--846, (2013).


\bibitem{10}
T. Barreiro, O. Bertolami and P. Torres, \textit{WMAP Five-Year Data Constraints on the Unified Model of Dark Energy and Dark Matter}, Phys. Rev. D, 78, 043530, (2008).

\bibitem{11_Nojiri_2006}
S. Nojiri and S. D. Odintsov, \textit{Unifying Phantom Inflation with Late-Time Acceleration: Scalar Phantom--Non-Phantom Transition Model and Generalized Holographic Dark Energy}, Gen. Relativ. Gravit., 38, 1285--1304, (2006).

\bibitem{12_Cai_2007}
R.-G. Cai, \textit{A Dark Energy Model Characterized by the Age of the Universe}, Phys. Lett. B, 657, 228--231, (2007).

\bibitem{13_Wei_2008}
H. Wei and R.-G. Cai, \textit{A New Model of Agegraphic Dark Energy}, Phys. Lett. B, 660, 113--117, (2008).

\bibitem{14_Wei_2008}
H. Wei and R.-G. Cai, \textit{Cosmological Constraints on New Agegraphic Dark Energy}, Phys. Lett. B, 663, 1--6, (2008).

\bibitem{17_Zhang_2013}
J.-F. Zhang, Y.-H. Li and X. Zhang, \textit{A Global Fit Study on the New Agegraphic Dark Energy Model}, Eur. Phys. J. C, 73, 2280, (2013).

\bibitem{ASSMUDAP25}
A. Sardar, S. Maity, U. Debnath and A. Pradhan, \textit{Different horizon cut-offs for Tsallis, Rényi and Sharma-Mittal holographic dark energies in Hořava-Lifshitz gravity}, Annals of Physics, Vol. 473, 169891 (2025). https://doi.org/10.1016/j.aop.2024.169891.

\bibitem{hh1} S. Maity, P. Rudra, \textit{Gravitational waves driven by holographic dark energy}, Nucl. Phys. B, 1009, 116724 (2024).


\bibitem{hh2} P. Saha, P. Rudra, \textit{A cosmological holographic reconstruction of $f(Q)$ theory}, Int. J. Mod. Phys. D., 34, 2550006 (2005).


\bibitem{hh3} M. Ghosh, P. Rudra, S. Chattopadhyay, B. Pourhassan, \textit{Warm inflation with Barrow holographic dark energy}, Nucl. Phys. B, 1017, 116933 (2025). 

\bibitem{hh4} P. Rudra, \textit{Ricci-Cubic holographic dark energy}, Phys. Dark Univ. 42, 101307 (2023).

\bibitem{hh5} S. Maity, P. Rudra, \textit{Inflation driven by Barrow holographic dark energy}, J. Hologr. Appl. Phys., 2, 1 (2022)

\bibitem{Neupane:2007ra}
I. P. Neupane, \textit{A Note on Agegraphic Dark Energy}, Phys. Lett. B, 673, 111--118, (2009).

\bibitem{SMASPR26}
S. Maity, A. Sanyal and P. Rudra, \textit{New agegraphic dark energy in loop quantum cosmology: a quantum gravitational perspective on dark energy evolution}, Eur. Phys. J. C 86, 211 (2026). https://doi.org/10.1140/epjc/s10052-026-15375-y

\bibitem{SMUD19}
S. Maity and U. Debnath, \textit{Tsallis, Rényi and Sharma-Mittal holographic and new agegraphic dark energy models in D-dimensional fractal universe}, Eur. Phys. J. Plus 134, 514 (2019). https://doi.org/10.1140/epjp/i2019-12884-6

\bibitem{PSSMUD22}
P. Saha, S. Maity and U. Debnath, \textit{Reconstructing extended f(P) cubic gravity from entropy-corrected holographic and new agegraphic dark energy models}, Mod. Phy. Let. A, Vol 37, 2250204 (2022). doi:10.1142/S0217732322502042.

\bibitem{SMAK25}
S. Maity and A. Kotal, \textit{Barrow agegraphic and new barrow agegraphic dark energy driven reconstruction of f(R) gravity and parameter constraints from observational data}, Phy. of Dark Univ., Vol. 50, 102184 (2025).

\bibitem{LI20041}
M. Li, \textit{A Model of Holographic Dark Energy}, Phys. Lett. B, 603, 1--5, (2004).

\bibitem{Wei:2007ty}
H. Wei and R.-G. Cai, \textit{A New Model of Agegraphic Dark Energy}, Phys. Lett. B, 660, 113--117, (2008).

\bibitem{21_PhysRevD.74.086009}
S. Nojiri, S. D. Odintsov and H. Štefančić, \textit{Transition from a Matter-Dominated Era to a Dark Energy Universe}, Phys. Rev. D, 74, 086009, (2006).

\bibitem{Zunckel2008}
C. Zunckel and C. Clarkson, \textit{Consistency Tests for the Cosmological Constant}, Phys. Rev. Lett., 101, 181301, (2008).

\bibitem{Sahni2008}
V. Sahni, A. Shafieloo and A. A. Starobinsky, \textit{Two New Diagnostics of Dark Energy}, Phys. Rev. D, 78, 103502, (2008).

\bibitem{Shafieloo2009}
A. Shafieloo, V. Sahni and A. A. Starobinsky, \textit{Is Cosmic Acceleration Slowing Down?}, Phys. Rev. D, 80, 101301, (2009).

\bibitem{lindd}
R. R. Caldwell and E. V. Linder, \textit{The Limits of Quintessence}, Phys. Rev. Lett. \textbf{95}, 141301 (2005).

\bibitem{Sahni:2002fz}
V. Sahni, T. D. Saini, A. A. Starobinsky and U. Alam, \textit{Statefinder: A New Geometrical Diagnostic of Dark Energy}, JETP Lett. \textbf{77}, 201--206 (2003).

\bibitem{Alam:2003sc}
U. Alam, V. Sahni, T. D. Saini and A. A. Starobinsky, \textit{Exploring the Expanding Universe and Dark Energy Using the Statefinder Diagnostic}, Mon. Not. Roy. Astron. Soc. \textbf{344}, 1057 (2003).

\bibitem{Jimenez2002}
R. Jimenez and A. Loeb, \textit{Constraining Cosmological Parameters Based on Relative Galaxy Ages}, Astrophys. J. \textbf{573}, 37--42 (2002).

\bibitem{Moresco2015}
M. Moresco, \textit{Raising the Bar: New Constraints on the Hubble Parameter with Cosmic Chronometers at $z\sim2$}, Mon. Not. Roy. Astron. Soc. Lett. \textbf{450}, L16--L20 (2015).

\bibitem{Moresco2012}
M. Moresco \textit{et al.}, \textit{Improved Constraints on the Expansion Rate of the Universe up to $z\sim1.1$ from the Spectroscopic Evolution of Cosmic Chronometers}, JCAP \textbf{08}, 006 (2012).

\bibitem{Moresco2016}
M. Moresco \textit{et al.}, \textit{A 6\% Measurement of the Hubble Parameter at $z\sim0.45$: Direct Evidence of the Epoch of Cosmic Re-Acceleration}, JCAP \textbf{05}, 014 (2016).





\bibitem{DESI2024}
DESI Collaboration, \textit{DESI 2024 III: Baryon Acoustic Oscillations from Galaxies and Quasars}, JCAP \textbf{04}, 012 (2025).

\bibitem{eBOSS2020}
eBOSS Collaboration \textit{et al.}, \textit{The Completed SDSS-IV Extended Baryon Oscillation Spectroscopic Survey: Cosmological Implications from Two Decades of Spectroscopic Surveys at the Apache Point Observatory}, Phys. Rev. D \textbf{103}, 083533 (2021).

\bibitem{DESI2025}
DESI Collaboration, \textit{DESI DR2 Results II: Measurements of Baryon Acoustic Oscillations and Cosmological Constraints}, Phys. Rev. D 112, 083515 (2025).

\bibitem{ForemanMackey2013}
D. Foreman-Mackey, D. W. Hogg, D. Lang and J. Goodman, \textit{emcee: The MCMC Hammer}, Publ. Astron. Soc. Pac. \textbf{125}, 306--312 (2013).

\bibitem{Lewis2019}
A. Lewis, \textit{GetDist: A Python Package for Analysing Monte Carlo Samples}, JCAP \textbf{08}, 025 (2025).

\bibitem{Planck2018}
N. Aghanim \textit{et al.} (Planck Collaboration), \textit{Planck 2018 Results. VI. Cosmological Parameters}, Astron. Astrophys. \textbf{641}, A6 (2020), arXiv:1807.06209 [astro-ph.CO].

\bibitem{Verde2019}
L. Verde, T. Treu and A. G. Riess, \textit{Tensions between the Early and Late Universe}, Nature Astronomy \textbf{3}, 891 (2019), arXiv:1907.10625 [astro-ph.CO].

\bibitem{Riess2022}
A. G. Riess \textit{et al.}, \textit{A Comprehensive Measurement of the Local Value of the Hubble Constant with 1 km/s/Mpc Uncertainty from the Hubble Space Telescope and the SH0ES Team}, Astrophys. J. Lett. \textbf{934}, L7 (2022), arXiv:2112.04510 [astro-ph.CO].

\bibitem{t1} J-X. Li, S. Wang, \textit{Revisiting the Hubble tension problem in the framework of holographic dark energy}, Mon. Not. Roy. Astron. Soc., \textbf{548}, 1 (2026)

\bibitem{t2} X. Tang \textit{et al.}, \textit{Constraining holographic dark energy and analyzing cosmological tensions}, Physics of the Dark Universe, \textbf{46}, 101568 (2024)

\bibitem{SanyalRudra2026}
A. Sanyal and P. Rudra, \textit{Ricci-cubic Holographic Dark Energy: Confronting Observations, Stability and the Cosmic Coincidence Problem}, Phys. Lett. B \textbf{878}, 140550 (2026).

\end{thebibliography}
\end{document}